\documentclass{raa_twocolumn}       
\usepackage{graphicx,times}
\usepackage{natbib}
\usepackage{amssymb,amsmath}
\usepackage{multirow}
\usepackage{multicol}
\usepackage{CJK}
\bibpunct{(}{)}{;}{a}{}{,}

\usepackage{hyperref}
\hypersetup{colorlinks = true, linkcolor = green, anchorcolor = red, citecolor = blue, filecolor = red, pagecolor = red, urlcolor = red}

\begin{document}
\begin{CJK*}{UTF8}{gbsn}

   \title{The Role of Magnetic Fields in Triggered Star Formation of RCW 120}

 \volnopage{ {\bf 2022} Vol.\ {\bf X} No. {\bf XX}, 000--000}
   \setcounter{page}{1}

   \author{Zhiwei Chen(陈志维)\inst{1}, Ramotholo Sefako \inst{2}, Yang Yang \inst{1,3}, Zhibo Jiang(江治波) \inst{1}, Shuling Yu(于书岭) \inst{1,3}, and Jia Yin(尹佳)\inst{4}}

   \institute{ Purple Mountain Observatory, Chinese Academy of Sciences, 10 Yuanhua Road, 210023 Nanjing, China; {\it zwchen@pmo.ac.cn}\\
   \and
       South African Astronomical Observatory, PO Box 9, Observatory 7935, Cape Town, South Africa
	\and
	 University of Science and Technology of China, Chinese Academy of Sciences, 230026 Hefei, China\\
       \and
       National Astronomical Observatories, Chinese Academy of Sciences, 100101 Beijing, China \\ 
         \vs \no
   {\small Received 2022 Feb 18; accepted 2022 xx xx}}

\abstract{We report on the near-IR polarimetric observations of RCW 120 with the 1.4 m IRSF telescope. The starlight polarization of the background stars reveals for the first time the magnetic field of RCW 120. The global magnetic field of RCW 120 is along the direction of $20\degr$, parallel to the Galactic plane. The field strength on the plane of the sky is $100\pm26\,\mu$G. The magnetic field around the eastern shell shows evidence of compression by the \ion{H}{ii} region. The external pressure (turbulent pressure + magnetic pressure) and the gas density of the ambient cloud are minimum along the direction where RCW 120 breaks out, which explains the observed elongation of RCW 120. The dynamical age of RCW 120, depending on the magnetic field strength, is $\sim\,1.6\,\mathrm{Myr}$ for field strength of $100\,\mu$G, older than the hydrodynamic estimates. In direction perpendicular to the magnetic field, the density contrast of the western shell is greatly reduced by the strong magnetic field. The strong magnetic field in general reduces the efficiency of triggered star formation, in comparison with the hydrodynamic estimates. Triggered star formation via the "collect and collapse" mechanism could occur in the direction along the magnetic field. Core formation efficiency (CFE) is found to be higher in the southern and eastern shells of RCW 120 than in the infrared dark cloud receiving little influence from the \ion{H}{ii} region, suggesting increase in the CFE related to triggering from ionization feedback.
}
\authorrunning{Zhiwei Chen at al.}
\titlerunning{Magnetic Fields of RCW 120}
\maketitle

\section{Introduction}           
\label{sect:intro}
An \ion{H}{ii} region is the result of the neutral gas ionized by the high-energy photons ($\geqslant13.6\,\mathrm{eV}$) emitted by massive stars ($M\geqslant8\,M_\odot$) \citep{1939ApJ....89..526S}. After the ignition of nucleosynthesis in a massive star, an \ion{H}{ii} region rapidly expands to the Str{\"o}mgren radius where the ionization rate is balanced with the recombination rate. The \ion{H}{ii} region now becomes a sphere with a thermal pressure exceeding the external pressure of the ambient interstellar medium (ISM), and a supersonic shock wave starts to propagate outwards into the surrounding ISM. A spherical shell of molecular gas swept-up by the shock is accumulated between the ionization front and shock front. Once the swept-up shell becomes dense enough to be gravitationally unstable, star formation in the swept-up shell could begin. This process, commonly called ``collect and collapse'' (C\&C), is the classical mechanism in which massive stars trigger next-generation star formation \citep{1977ApJ...214..725E,1994MNRAS.268..291W}. Another mechanism, radiation-driven implosion (RDI), for the triggered star formation due to the expansion of an \ion{H}{ii} region describes the interaction between a pre-existing, stable clump and the \ion{H}{ii} region that accelerates the star formation of the stable clump \citep{1989ApJ...346..735B,2003MNRAS.338..545K}. Because in reality the ISM is not uniform, and the internal structure is fractal, it is always not straightforward to classify the exact mechanisms (C\&C and/or RDI) responsible for the star formation triggered by an \ion{H}{ii} region. Recently, a hybrid mechanism consisting of both C\&C and RDI was proposed to explain the existence of dense cores and star formation on the swept-up shell, based on the hydrodynamic simulations of an \ion{H}{ii} region expanding into the fractal molecular cloud \citep{2015MNRAS.452.2794W}. 

The star formation triggered by an \ion{H}{ii} region has attracted intensive interests since infrared (IR) bubbles were ubiquitously found in the Galactic disk \citep{2006ApJ...649..759C,2007ApJ...670..428C,2015ApJ...799..153K,2019MNRAS.488.1141J}. The majority of IR bubbles are a type of object created by the expansion of an \ion{H}{ii} region, and are characterized by the (partially) bound photodissociation region (PDR) between the ionized region and the surrounding molecular cloud. IR bubbles are ideal targets for investigating the possible relation between the star formation occurring on the swept-up shell and the ionization feedback. The C\&C and RDI mechanisms are both proposed to explain the observed star formation on the swept-up shells of IR bubbles \citep[e.g.,][]{2010A&A...523A...6D,2010ApJ...716.1478W,2015MNRAS.450.1199D}. Although the consensus on the star formation triggered by an \ion{H}{ii} region is established based on the hydrodynamic simulations and observational studies, the role of the magnetic field in this process is far from well understood. \citet{1989ApJ...346..735B} discussed the role of the magnetic field in the RDI process for a spherical cloud, and found that the density contrast of the swept-up shell depends mostly on the strength of the magnetic fields in the initial cloud. Magnetohydrodynamic (MHD) simulations of the evolution of an \ion{H}{ii} region into the uniform magnetized ISM showed that the expansion velocity of the \ion{H}{ii} region is fastest along the magnetic fields in the initial ISM, and is slowest perpendicular to the magnetic fields. This leads to the formation of an \ion{H}{ii} region that is bounded by a dense shell of swept-up gas in the direction along the magnetic field, but not perpendicular to it \citep{2007ApJ...671..518K,2011MNRAS.414.1747A}. These MHD simulations also predict a magnetically critical radius at which the thermal pressure of the \ion{H}{ii} region is comparable to the pressure of magnetic fields. When the \ion{H}{ii} region expands to a radius larger than this magnetically critical radius, the effects of magnetic fields on the \ion{H}{ii} region become important, and the most obvious result is that magnetic fields deform the originally spherical \ion{H}{ii} region to be elongated along the magnetic fields. The ordered magnetic fields prevent the \ion{H}{ii} region from collecting and compressing as much gas as one might expect from a purely hydrodynamic estimate \citep{2007ApJ...671..518K}, and also reduce fragmentation in the virialized cloud \citep{2015MNRAS.454.4484G}. Therefore magnetic fields may reduce the efficiency of triggered star formation from an \ion{H}{ii} region \citep{2007ApJ...671..518K}. These MHD simulations do not include self-gravity, and thus are unable to model triggered star formation on the swept-up shell of \ion{H}{ii} region. A few observational results on the magnetic fields of \ion{H}{ii} regions attempt to address the relation between magnetic fields and triggered star formation \citep{2017ApJ...838...80C,2018ApJ...869L...5L,2021ApJ...911...81D}. However these observational results manifest great diversity. To date there have been far fewer systematic studies on the interaction of \ion{H}{ii} region feedback with magnetic fields than for outflow feedback, and thus the range of possible effects is quite uncertain \citep{2019FrASS...6....7K}.

RCW 120 is a well-studied Galactic \ion{H}{ii} region because of its ovoid shape and its relatively close distance at 1.34 kpc \citep{2007A&A...472..835Z} or at 1.68 kpc \citep{2019ApJ...870...32K}. In this paper, we adopt the most recent distance estimate of 1.68 kpc. RCW 120 is the testbed for studying the interaction of \ion{H}{ii} region feedback with the ambient gas, dust and triggered star formation \citep{2007A&A...472..835Z,2010A&A...518L..81Z,2020A&A...638A...7Z,2009A&A...496..177D,2010A&A...510A..32M,2010A&A...518L..99A,2012A&A...542A..10A,2015ApJ...800..101A,2013ARep...57..573P,2014A&A...564A.106T,2014A&A...566A..75O,2015ApJ...806....7T,2015A&A...579A..10R,2015MNRAS.452.2794W,2016A&A...586A.114M,2017MNRAS.469..630A,2017A&A...600A..93F,2018A&A...616L..10F,2020A&A...639A..93F,2019MNRAS.483..352M,2019A&A...631A.170R,2019MNRAS.488.5641K,2021MNRAS.503..633K,2021SciA....7.9511L,2022A&A...659A..36K}. These results greatly enrich our understanding of RCW 120, especially regarding star formation on the boarder of RCW 120. \citet{2011MNRAS.414.1747A} studied the evolution of RCW 120 as it expands into the magnetized ISM with uniform density. In these MHD simulations with magnetic field strength of $24.16\,\mu$G, the role of magnetic fields is minor compared to the thermal pressure of RCW 120 within the limited simulation time. However, the role of magnetic fields depends strongly on the field strength. The observed morphology of RCW 120 is much more elongated than the simulated morphology in \citet{2011MNRAS.414.1747A}, implying that a stronger magnetic field is needed to explain the elongation of RCW 120. In this paper, we use the near-IR starlight polarization to reveal, for the first time, the magnetic fields of RCW 120.


\section{Data Acquisition}
\label{sect:Obs}
\subsection{IRSF/SIRPOL Observations}
The data were taken with SIRPOL on the 1.4 m InfraRed Survey Facility (IRSF) telescope at the South African Astronomical Observatory, in Sutherland, South Africa. SIRPOL is a single-beam polarimeter with an achromatic half-wave plate rotator unit and a polarizer attached to the near-IR camera SIRIUS \citep{2003SPIE.4841..459N,2006SPIE.6269E..51K}. SIRPOL enables wide-field ($\sim 8\arcmin\times8\arcmin$) polarization imaging with pixel scale of $0\farcs45$ per pixel in the $JHK_s$ bands simultaneously. The observations were made in the night of 2018 July 20. To cover the entire extent of RCW 120, we designed a grid of six positions on the sky. In each grid position, two sets of observations were conducted. Each set took 30\,s exposures at four wave-plate angles (in the sequence of $0\degr$, $45\degr$, $22\fdg5$, and $67\fdg5$) at 10 dithered positions. The total integration time was 600\,s per wave plate angle in one grid position. Including observation overheads, the total observation time in one grid position was 54 mins. The total time spent on the entire extent of RCW 120 was 5.7 hr. The airmass during the observations  was in the range of $1.032-1.722$. Observations with lower airmass generally have smaller full width at half maximum (FWHM) values. The measured FWHM averaged for the point sources in the six grid positions has a range from less than 3 pixels to close to 4 pixels, or $\sim1\farcs5-1\farcs8$.

The data were processed using the pyIRSF pipeline (version 3.0) developed by Y. Nakajima \footnote[1]{The tarball file of pyIRSF is available at https://sourceforge.net/projects/irsfsoftware}, including dark-field subtraction, flat-field correction, median sky subtraction and frame registration. The final products of pyIRSF are the $JHK_s$ scientific images at four wave plate angles in six grid positions. 
  
The FWHM in different grid positions can vary by more than one pixel. In each grid position, we computed $\langle \mathrm{FWHM} \rangle$ for the point sources in each of the $JHK_s$ bands. A fixed aperture of 4 pixels slightly larger than the corresponding $\langle \mathrm{FWHM} \rangle$ was used in the aperture photometry for the $JHK_s$ images of the six grid positions. We utilized the aperture photometry routine \textsl{sirphot.py} embedded in pyIRSF to extract the digital counts of point sources in the $JHK_s$ bands. The \textsl{sirphot.py} routine in pyIRSF is written with pyRAF under the Python 2.7 environment. The output catalog of \textsl{sirphot.py} contains pixel coordinates, digital counts and uncertainties, FWHM and ellipticity of the point sources with peak intensities higher than the specified threshold, $5\times \sigma_\mathrm{sky} + I_\mathrm{sky}$, where $\sigma_\mathrm{sky}$ and $I_\mathrm{sky}$ are the median noise and intensity of sky background respectively. The FWHM and ellipticity help to exclude some point sources that are blended with nearby sources, because aperture photometry of fixed aperture radius for these blended sources most likely includes contamination from nearby sources. Therefore, we only keep point sources with FWHM smaller than 5 pixels and with ellipticity smaller than 0.3. These criteria could exclude most of the blended sources. 
 

For every source detected at the four wave plate angles, the Stokes parameters $I$, $U$, and $Q$ are computed by the following formulas:
\begin{align}
 I & =  (I_0+I_{22.5}+I_{45}+I_{67.5})/2 \\
 Q & =  (I_0-I_{45})/I \\
 U & =  (I_{22.5}-I_{67.5})/I
\end{align}
The source counts at the four wave plate angles are $I_0$, $I_{22.5}$, $I_{45}$, and $I_{67.5}$, which are directly obtained from the aperture photometry.

Some instrument calibrations must be taken into account in the conversion from the observed Stokes parameters $I$, $U$, and $Q$ to the polarization degree and position angle (P.A.). The polarization efficiency $\eta$ of SIRPOL is 95.5\%, 96.3\%, and 98.5\%, and the correction angle $\theta_0$ is $105\degr$ in each of the $JHK_s$ bands, respectively \citep{2006SPIE.6269E..51K}. The equatorial Stokes $Q_\mathrm{eq}$ and $U_\mathrm{eq}$ values are computed as follows \citep{2018PASP..130e5002D}:
\begin{align}
  Q_\mathrm{eq}& = (Q \cos(2\theta_0) - U\sin(2\theta_0))/\eta\\
  U_\mathrm{eq}& = (U\cos(2\theta_0) + Q\sin(2\theta_0))/\eta
\end{align}

Furthermore, $Q_\mathrm{eq}$ and $U_\mathrm{eq}$ are combined to form the equatorial degree of polarization $P_\mathrm{eq}$, and the P.A. $\theta_\mathrm{PA}$, counted counterclockwise from the north to the east, which are given by Equations (6) and (7).   
\begin{align}
P_\mathrm{eq}& = \sqrt{U_\mathrm{eq}^2+Q_\mathrm{eq}^2} \\
\theta_\mathrm{PA} & = \frac{1}{2}\,\arctan(U_\mathrm{eq}/Q_\mathrm{eq})
\end{align}

The polarization degree uncertainty $\delta P$ is computed from the corresponding uncertainties of Stokes $I$, $Q$, and $U$ \citep[Appendix B]{2018PASP..130e5002D}. The debiased polarization degree $P$ is obtained according to the method \citep{1974ApJ...194..249W}
\begin{align}
P = \sqrt{P^2_\mathrm{eq} - \delta P^2}
\end{align}
The uncertainty of $\theta_\mathrm{PA}$ is then computed by $\delta \theta_\mathrm{PA} = 28\fdg6 \delta P/P$; thus smaller $\delta P/P$ corresponds to smaller $\delta \theta_\mathrm{PA}$. In the following analysis, we only consider point sources satisfying $P\geqslant 2\,\delta P$.

\subsection{Infrared Catalog of the Southern Galactic Plane}

The Vista Variables in the V{\'i}a L{\'a}ctea (hereafter VVV) \citep{2010NewA...15..433M} is a European Southern Observatory (ESO) public near-IR survey that covers approximately 562 deg$^2$ areas of the inner Galactic plane with VIRCAM on the 4m VISTA telescope. The VVV survey imaged the inner Galactic plane in the $ZYJHK_s$ bands. RCW 120 is covered by the VVV survey, which provides deep near-IR images for this \ion{H}{ii} region. Two point spread function (PSF) photometric catalogs for the VVV survey have been released, one by the VVV science team \citep{2018A&A...619A...4A} and another one by personal contribution \citep[hereafter ZK19]{2019A&A...632A..85Z}, which reaches on average about one magnitude deeper than the former. Considering this advantage, the ZK19 catalog is utilized in this paper. The ZK19 catalog applies the Vega photometric system in the $JHK_s$ bands, and the $JHK_s$ magnitudes match the 2MASS system.

\begin{figure*} 
   \centering
   \includegraphics[width=0.9\textwidth]{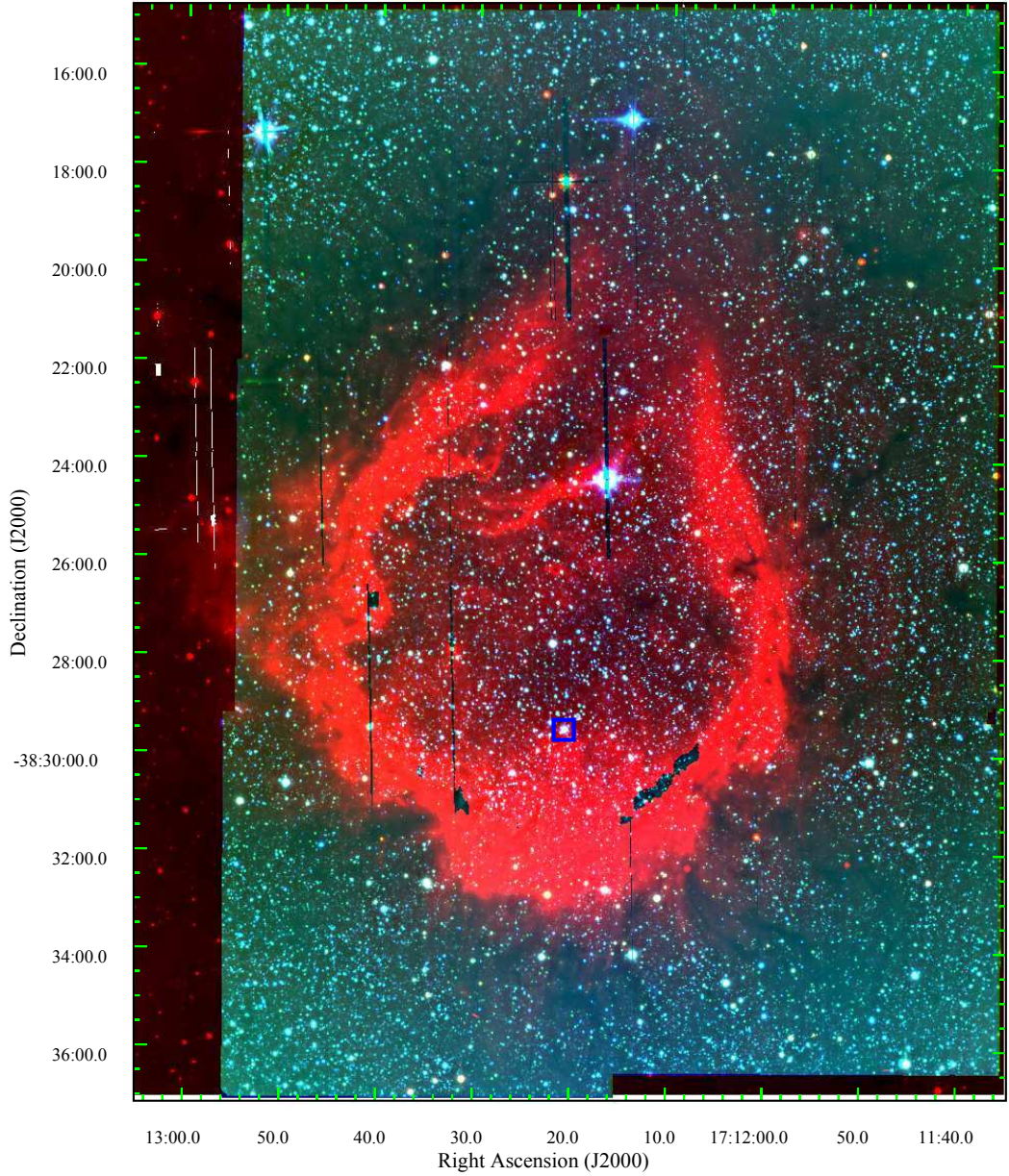}
   \caption{Three-color image of RCW 120, made from the SIRPOL $H$ (blue), SIRPOL $K_s$ (green) and Spitzer $8.0\,\mu$m (red) images. The ionizing source of RCW 120 is marked by the blue square closer to the southern shell.} 
   \label{Fig:HKI4}
 \end{figure*}
 
\subsection{CO Molecular Line Data of RCW 120}\label{Sect:CO}
The CO and CO isotopes $J=1-0$ data were previously used by \citet{2015ApJ...800..101A} to study the molecular gas environment of RCW 120. The $^{12}$CO, $^{13}$CO, and C$^{18}$O $J=1-0$ data were taken from 2011 July 15 to 18 with the Australian National Telescope Facility (ANTF) Mopra 22 m radio telescope in New South Wales, Australia. See more details of the Mopra observations in \citet{2015ApJ...800..101A}. We retrieved the raw data from Australia Telescope Online Archive, and reduced the raw data using the ANTF GRIDZILLA \citep{1995ASPC...77..433S} and LIVEDATA \citep{2001MNRAS.322..486B} packages with the same configurations as \citet{2015ApJ...800..101A}.

Based on the analyses of the CO molecular line emissions of RCW 120 in \citet{2015ApJ...800..101A}, the molecular line emissions from $-15$ to $-0.5 \,\mathrm{km\,s^{-1}}$ are used to derive the physical properties of the molecular gas associated with RCW 120. From the data cube of the CO, $^{13}$CO and C$^{18}$O emissions, we derive the moment 0 (integrated intensity) and line width maps of the three lines. 

The optically thick line $^{12}$CO is utilized to derive the excitation temperature $T_\mathrm{ex}$ by the formula
\begin{align}
T_\mathrm{ex}(^{12}\mathrm{CO})& = \frac{h\nu}{k}\left[\ln \left(1 + \frac{h\nu/k}{T_\mathrm{MB}+ J_\nu(T_\mathrm{bg})}\right)\right]^{-1},
\end{align}
where $J_\nu(T_\mathrm{bg}) = \frac{h\nu/k}{\exp{(h\nu/k T_\mathrm{bg})} -1}$ is the blackbody radiation of the cosmic background of $T_\mathrm{bg}=2.73\,\mathrm{K}$ at the frequency $\nu=115.2712\,\mathrm{GHz}$ of the $^{12}$CO $1-0$ line. Assuming that the molecular gas is in the state of local thermal equilibrium (LTE), the CO and its isotopes are sharing the same excitation temperature $T_\mathrm{ex}$ of $^{12}$CO. 
The $^{13}$CO $1-0$ line emission is mostly optically thin, and becomes optically thick in very dense regions. The C$^{18}$O $1-0$ line emission, on the other hand, is ubiquitously found to be optically thin. The two molecules are able to be representative of the molecular gas with high density. In the state of LTE, the column density of $^{13}$CO and C$^{18}$O then can be calculated from the $T_\mathrm{ex}$ and the main beam brightness temperature of the corresponding line emission, that is \citep{2016PASP..128b9201M}
\begin{align}\label{Equ:N}
N_\mathrm{tot} & = \frac{2.48\times10^{14} (T_\mathrm{ex} + 0.88)\exp{(E_u/T_\mathrm{ex})}}{\exp{(E_u/T_\mathrm{ex})-1}}\times \nonumber\\
              & \frac{\int T_\mathrm{MB} dv (\mathrm{km\,s^{-1}})}{f(J_\nu(T_\mathrm{ex}) - J_\nu(T_\mathrm{bg}))} \,\mathrm{cm^{-2}},
\end{align}
where the filling factor $f=1$, and $J_\nu(T_\mathrm{ex})$ is the blackbody radiation of the excitation temperature at the frequency of the corresponding line. $E_u = 5.29\,\mathrm{K}$ for $^{13}$CO $1-0$, and $E_u = 5.27\,\mathrm{K}$ for C$^{18}$O $1-0$. The optical depth of the $^{13}$CO $1-0$ line emission can be $\gtrsim1$ in very dense regions. The $\tau_{13}$ of the $^{13}$CO $1-0$ line emission is solved by
\begin{align}\label{Equ:tao}
\tau_\mathrm{13} & = -\ln\left[ 1- \frac{T_\mathrm{MB}}{f(J_\nu(T_\mathrm{ex}) - J_\nu(T_\mathrm{bg}))}\right],
\end{align}
and the optical depth correction factor $\frac{\tau_{13}}{1-\exp{(-\tau_{13})}}$ is multiplied by the column density of $^{13}$CO in Equation~\ref{Equ:N} to obtain a better estimate of $N(\mathrm{^{13}CO})$. For the C$^{18}$O $1-0$ line emission, the optical depth is always $\ll1$, and the optical depth correction factor is $\approx1$. The map of $N(\mathrm{^{13}CO})$ and $N(\mathrm{C^{18}O})$ is generated for the emission from $-15$ to $-0.5 \,\mathrm{km\,s^{-1}}$ and is smoothed with a Gaussian filter of FWHM 3 pixels. The mean $^{13}$CO to H$_2$ abundance is $2\times10^{-6}$ \citep{1978ApJS...37..407D}. The $^{13}$CO to C$^{18}$O abundance is set to $\sim7$, an averaged value for the $A_V$ range of $3-15$ mag \citep{2004A&A...421.1087H}. From the map of $N(\mathrm{^{13}CO})$ and $N(\mathrm{C^{18}O})$, we can obtain the map of molecular hydrogen column density $N(\mathrm{H_2})$ by the ratios $N(\mathrm{H_2})/N(\mathrm{^{13}CO})=5\times10^5$ and $N(\mathrm{H_2})/N(C\mathrm{^{18}O})=3.5\times10^6$. The CO data provide the fundamental properties of the molecular gas associated with RCW 120, which are vital for understanding the evolution of RCW 120.


\section{Results}\label{Sect:res}
Figure~\ref{Fig:HKI4} displays the three-color composite image of RCW 120. The combined field-of-view (FOV) of the near-IR polarization observations is $15\farcm6\times22\farcm1$, and covers the entire RCW 120 \ion{H}{ii} region and the shell of neutral gas, and substantial part of the surrounding molecular clouds. Only a small portion of the $8\,\mu$m emission extending to the far east is not observed with IRSF/SIRPOL. This arc-like structure to the east of the shell seen at $8\,\mu$m is probably the local PDR created by the two Herbig Ae/Be objects 1 and 2 \citep[see Figure 8 in][]{2007A&A...472..835Z}. These are not massive enough to form \ion{H}{ii} regions, but they can heat the surrounding dust and create local PDRs with lower energy photons. 

The near-IR polarization observations by SIRPOL are able to capture the magnetic fields within the \ion{H}{ii} region and the surrounding molecular clouds. Given the well-established relation between starlight polarization by dust extinction and interstellar magnetic fields \citep{2015ARA&A..53..501A}, the polarized starlight along the same line-of-sight (LOS) traces the orientations of magnetic fields on the plane of the sky (hereafter $B_\mathrm{sky}$). \citet{2012PASJ...64...110C} used SIRPOL to obtain $JHK_s$ polarimetric images of the giant \ion{H}{ii} region M17. The stars with polarization in the $JHK_s$ bands show ratios of polarization degree ($P_{\lambda1}/P_{\lambda2} \propto (\lambda1/\lambda2)^{-1.8}$) independent of colors, indicating the dichroic extinction origin for the polarization in the $JHK_s$ bands. The $B_\mathrm{sky}$ of M17 traced by these stars is consistent with the optical polarization in the \ion{H}{ii} region of low extinction and the far-IR dust emission polarization in the surrounding PDR with moderate extinction \citep{2012PASJ...64...110C}.

\subsection{Classifying Reddened Background Stars}\label{Sect:bkg}
The SIRPOL $JHK_s$ images are not sufficiently deep to penetrate through the dense PDR and molecular gas in RCW 120. In order to circumvent this problem, we combine the $JHK_s$ polarimetric catalog ($P_\lambda \geqslant 2 \delta P_\lambda$) generated by SIRPOL with the ZK19 $JHK_s$ photometric catalog. Applying a maximum separation of $1\arcsec$ in cross-match, we derive three polarimetric catalogs also with reliable $JHK_s$ photometry from the ZK19 catalog. There are 949 sources in the SIRPOL $J$ band, 1410 sources in the SIRPOL $H$ band and 1215 sources in the SIRPOL $K_s$ band, that are in common with the ZK19 $JHK_s$ photometry.

\begin{figure*} 
   \centering
   \includegraphics[width=\textwidth]{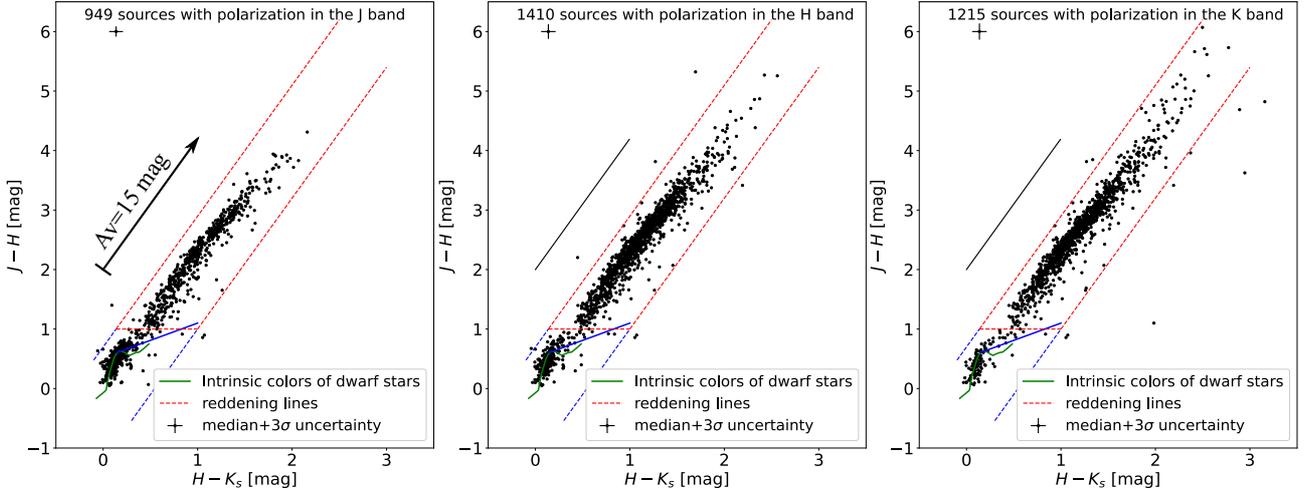}
   \caption{$J-H$ vs. $H-K_s$ color-color diagrams of the sources in the $JHK_s$ bands from left to right. The main-sequence intrinsic colors (highlighted in green) are adopted from \citet{2013ApJS..208....9P}. The intrinsic colors of T Tauri stars are displayed as the blue line \citep{1997AJ....114..288M}. } 
   \label{Fig:ccd}
   \end{figure*}
With the deep $JHK_s$ photometry by ZK19, we are able to apply the classical $J-H$ versus $H-K_s$ color-color diagram in classifying background stars that mostly show large $H-K_s$ and $J-H$ colors. Figure~\ref{Fig:ccd} features the color-color diagram for the polarimetric catalogs cross-matched with the ZK19 catalog in the $JHK_s$ bands. It is clear that two populations exist in the three color-color diagrams, stars with low $H-K_s$ and $J-H$ colors and others with higher colors and distributed along the reddening lines. The population with low $H-K_s$ and $J-H$ colors shows similar distributions with those of the main-sequence stars (the green curves in Figure~\ref{Fig:ccd}). If the main-sequence stars are reddened by a small amount of extinction ($A_V\sim0.8$) along the reddening lines, the distributions of main-sequence stars match better with those of the population with low $H-K_s$ and $J-H$ colors. We regard the population with low IR colors as the candidates for foreground stars which are lying in front of RCW 120. Vice versa, the population with higher color and distributed along the reddening lines is regarded as the candidates for stars whose large IR colors arise (in part) from the dust component of the local ISM in RCW 120. The criteria for background reddened stars are (vice versa for foreground stars):
\begin{align}
  J-H & \leqslant 2.2 \times (H-Ks) + 0.598 \\
  J-H & \geqslant 2.2\times (H-Ks) - 0.615 \\
  J-H & \geqslant 0.9 , 
\end{align}
and are shown as the red dashed lines in Figure~\ref{Fig:ccd}. The numbers of candidate foreground stars are 366, 222 and 106 in the $JHK_s$ bands, respectively. The numbers of candidate background stars are 565, 1159 and 1082 in the $JHK_s$ bands, respectively. 

\begin{figure*} 
   \centering
   \includegraphics[width=0.7\textwidth]{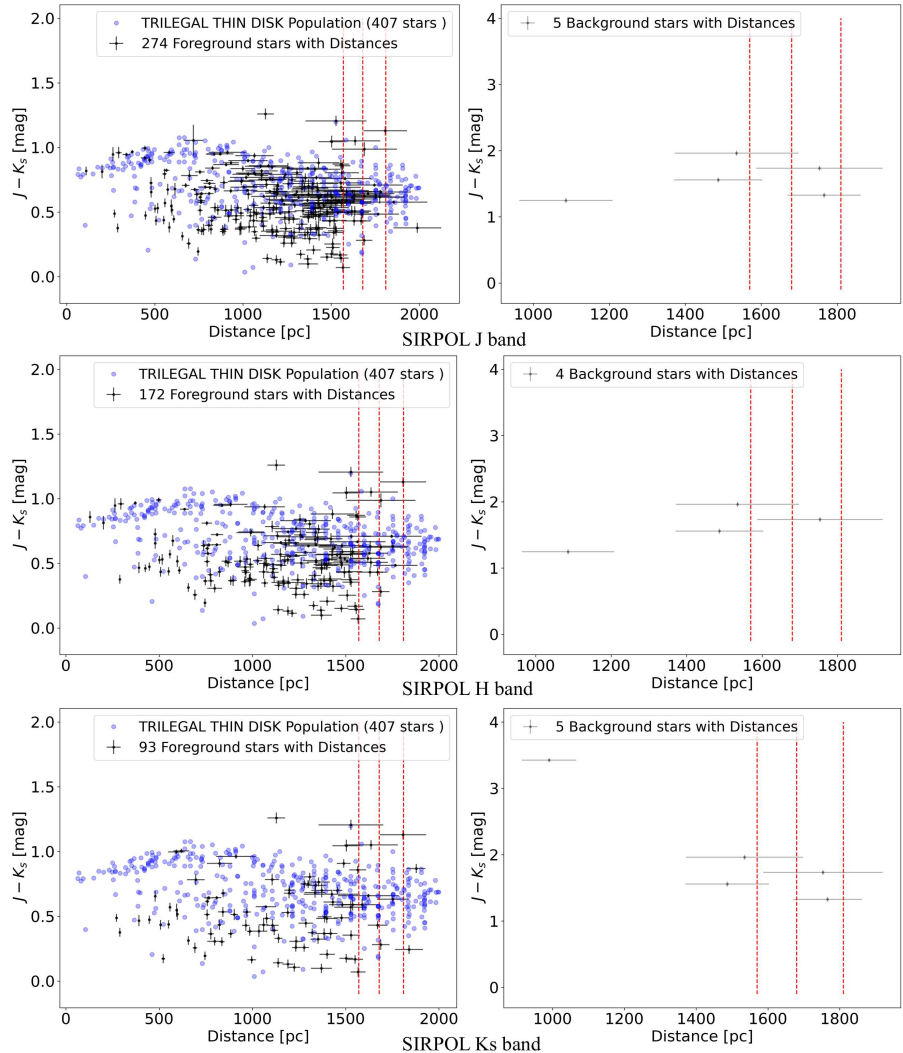}
   \caption{$J-K_s$ vs. distance diagram of the sources with accurate distance estimates (black dots with error bars) in the $JHK_s$ bands from top to bottom. The synthesized stellar populations from TRILEGAL are displayed as blue solid circles.} 
   \label{Fig:color_dist}
   \end{figure*}

The distance to RCW 120 is determined as $1680^{+130}_{-110}$\,pc \citep{2019ApJ...870...32K}. Aided by the accurate astrometric catalog from the Gaia Early Data Release 3 (EDR3) \citep{2016A&A...595A...1G,2021A&A...649A...1G}, a fraction of stars along the LOS toward RCW 120 has measured parallax values, which can be converted into distance estimates \citep{2021AJ....161..147B}. From the star distance catalog \citep{2021AJ....161..147B}, we retrieved distance estimates of the sources in the SIRPOL $JHK_s$ catalogs by using a match radius of $1\arcsec$. We considered the sources with a narrow range of distance estimates, defined by $(B_{rpgeo} - b_{rpgeo})/rpgeo < 0.2$, where $B_{rpgeo}$, $b_{rpgeo}$ and $rpgeo$ are the 84th percentile, the 16th percentile and the median value of the photogeometric distance posterior respectively \citep{2021AJ....161..147B}. 

Figure~\ref{Fig:color_dist} presents the diagram of ZK19 $J-K_s$ colors versus distances for the candidate foreground and background stars in the $JHK_s$ bands. For the candidate background stars with distances, it is straightforward to kick out the sources with $rpgeo < 1570$ pc. Among the thousands of candidate background stars, only a few stars have narrow distance estimates in the $JHK_s$ bands. The extremely low common fraction between candidate background stars and those with distances suggests that most of the candidate background stars have large extinction. This large extinction is also evidenced by the long path along the reddening lines of length about $A_V=35$. We regard all candidate background stars, except for a few with incompatible distances, as background stars with respect to RCW 120. The numbers of background stars are 562, 1156 and 1079 in the $JHK_s$ bands, respectively.

Most of the candidate foreground stars in the $JHK_s$ bands have narrow distance estimates. The situation for the candidate foreground stars is complicated, because not all candidate foreground stars have narrow distance estimates. In addition to the distributions of candidate foreground stars in Figure~\ref{Fig:color_dist}, we also show the distributions of synthesized stellar populations of the thin Galactic disk generated by TRILEGAL v1.6 \citep{2016AN....337..871G}. The synthesized stellar populations are displayed in blue dots in Figure~\ref{Fig:color_dist}. 


At first glance, the distributions of candidate foreground stars with distances match well with those of the synthesized stellar populations in Figure~\ref{Fig:color_dist}. For $0 < distance < 750$ pc and $0.7 < J-Ks < 1.2$, the stellar populations are dominated by the low-mass stars later than mid-K type \citep{2013ApJS..208....9P}, which are reproduced well by the TRILEGAL stellar populations.
As the distance increases, the low-mass stars later than mid-K type go to the very faint end which escape detection. Therefore, stars earlier than K-type (bluer in the $J-K_s$ colors) persist at the larger distances. As the distance increases, the interstellar extinction accumulated along the LOS starts to take effect in reddening the stars at larger distances. Both the effects from distance dilution and accumulated interstellar extinction merge to produce the dominant populations with $1000<distance<1500$\,pc and $0.5<J-K_s<0.8$. A small fraction of candidate foreground stars has distances comparable to or larger than that to RCW 120. For example, using the criterion $rpgeo > 1570$\,pc, 35 sources from the 520 candidate foreground stars in the $J$ band, 22 sources from the 410 candidate foreground stars in the $H$ band and 14 sources from the 334 candidate foreground stars in the $K_s$ band satisfy this criterion. In contrast, among the 407 stars generated by TRILEGAL, 105 stars are within the range $1570<distance<2000$, leading to a fraction of 28\% that is about 3 times the observed value (about 10\%). The TRILEGAL stars are generated under the universal relation $A_V=1.7$ mag kpc$^{-1}$ between interstellar extinction and distance. However, the molecular gas and interstellar dust associated with RCW 120 cause a sudden increase in extinction, which is responsible for blocking out most of the stars lying behind RCW 120. The dust extinction in RCW 120 is especially significant for the optical bandpass of Gaia. This can explain the very low fraction of candidate foreground stars with distances comparable to or larger than that to RCW 120. We examined the spatial distributions of these small number of candidate foreground stars. Most of them are located outside the bubble region, and most of them are within crowded fields that have little impact from extinction from the local dust in RCW 120. Considering the distance uncertainty of both RCW 120 and this minor sample of candidate foreground stars, we regard candidate foreground stars with $rpgeo\gtrsim1570$\,pc as background stars, resulting in 29, 14, and 9 more background stars in the $JHK_s$ bands, respectively. 

\begin{figure*}[t]
   \centering
   \includegraphics[width=\textwidth]{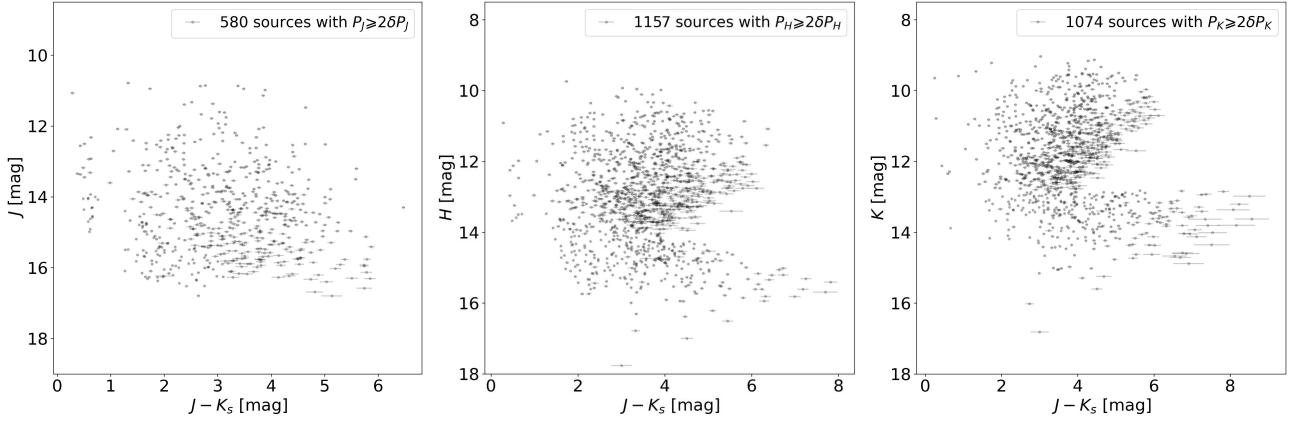}
   \caption{Magnitude vs. $J-K_s$ color diagram of the background sources in the $JHK_s$ bands from left to right respectively.} 
   \label{Fig:cmd}
   \end{figure*}

The above analyses for the distributions of both the observed stars and synthesized TRILEGAL stellar populations indicate that the bluer foreground stars overall consist of foreground stars with respect to RCW 120. A small fraction (a few percent) of candidate foreground stars at distance similar to RCW 120 do not substantially increase the number of background stars (a few tens compared to a thousand). We believe that the criteria used in the color-color diagram are reliable to distinguish largely reddened background stars and much bluer foreground stars.

Furthermore, we exclude possible young stellar objects (YSOs) which have $J-H$ and $H-K_s$ colors that resemble those of reddened diskless stars. We cross-match the reddened background stars in the $JHK_s$ bands with the Spitzer/IRAC candidate YSO catalog \citep{2021ApJS..254...33K} with a radius of $2\arcsec$. Eleven sources in the $J$ band, 13 sources in the $H$ band and 14 sources in the $K_s$ band are candidate YSOs. These candidate YSOs among the reddened background stars are excluded in further analyses. The above analyses yield 580 sources in the $J$ band, 1157 sources in the $H$ band and 1074 sources in the $K_s$ band, that are reddened background stars with respect to RCW 120.

\subsection{Polarization of reddened background stars}\label{Sect:cmd}
Reddened background stars are suffering extinction due to dust grains in RCW 120 and those behind RCW 120. Figure~\ref{Fig:cmd} plots the distributions of reddened background stars with $JHK_s$ polarization in the color-magnitude diagrams. The distributions of sources in the $HK_s$ bands clearly show evidence of the existence of distant background stars, defined by the regions $J-K_s > 4$ and $H<14$ or $K_s<12.8$. The stars in these regions manifest a trend of increase in luminosity as the $J-K_s$ colors increase. This trend is reversed compared to the normal trend that stars dim with increasing extinction. The most probable explanation for this reversed trend is that the stars in the regions are dominated by late-type (super) giants which are very luminous in the $HK_s$ bands. The late-type (super) giants are so luminous that they can be detected at far distances \citep{2013A&A...557A..51C}. 

The CO data of RCW 120 tell us that the molecular gas associated with RCW 120 is the dominant component along the LOS from near to far. The interstellar dust behind RCW 120 donates minor contribution to the total extinction of distant background stars. The local dust in RCW 120 plays the dominant role in producing the polarization of these stars. The starlight polarization due to dust extinction shows $\theta_\mathrm{PA}$ parallel to the $B_\mathrm{sky}$. The polarimetric catalogs in the $JHK_s$ bands are believed to contain background stars mostly reddened by the dust grains in RCW 120, and thus are capable of tracing the local $B_\mathrm{sky}$ of RCW 120. 

\begin{figure*} 
   \centering
   \includegraphics[width=0.95\textwidth]{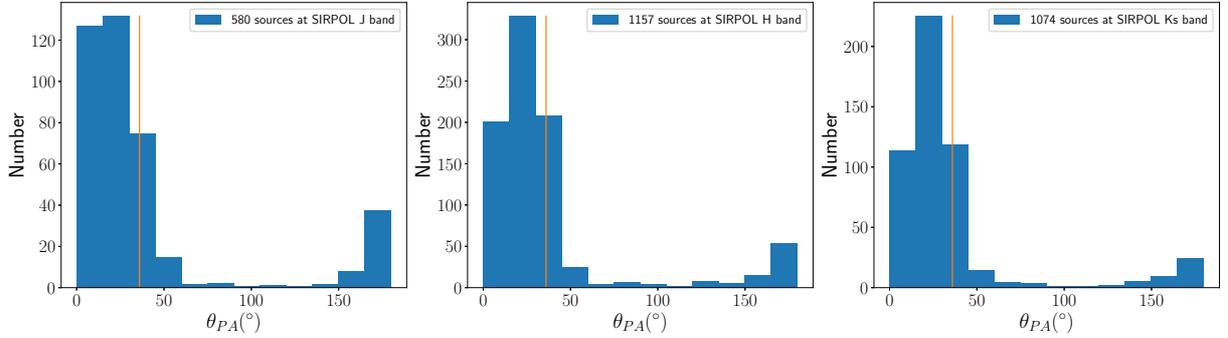}
   \caption{Histograms of $\theta_\mathrm{PA}$ of the starlight polarization sources in the $JHK_s$ bands from left to right respectively. The orientation of the Galactic plane is shown as the orange line in the plots.} 
   \label{Fig:all_PA}
   \end{figure*}

For a direct view on the orientations of the $JHK_s$ polarization, Figure~\ref{Fig:all_PA} presents the histograms of $\theta_\mathrm{PA}$ in the $JHK_s$ bands. The majority of background stars have $\theta_\mathrm{PA}$ values between $-15\degr$ ($165\degr$) and $45\degr$. The peak location is between $15\degr$ and $30\degr$. The distributions of $\theta_\mathrm{PA}$ in the $JHK_s$ are almost the same. The number of starlight polarization is highest in the $H$ band, in which the extinction is much smaller than in the $J$ band and the polarization degree is much higher than in the $K_s$ band. The $H$-band polarization is widely utilized to study the interstellar magnetic fields \citep{2012ApJS..200...19C,2019ApJ...875L..16T}. In the following, we use the $H$-band polarization of 1157 background reddened stars to trace the local $B_\mathrm{sky}$ of RCW 120. The $H$-band polarization data of these 1157 background stars are listed in Table~\ref{tbl:Htable}.

\begin{table*}
  \centering
  \caption{$H$-band Starlight Polarization of 1157 Background Stars }{\label{tbl:Htable}}
  \begin{tabular}{ccccccccccccc}
  
\hline\hline\noalign{\smallskip}
ID & R.A. & Decl. & $P_H$ & $\delta P_H$ & $\theta_{PA}$ & $\delta \theta_{PA}$ & $J$ & Jerr & $H$ & Herr & $K$ & Kerr \\
   & (deg)   & (deg)   & (\%)     & (\%)    & ($\degr$) & ($\degr$) & (mag)  &  (mag)& (mag)  & (mag) & (mag)  & (mag) \\
\hline
1 & 258.198666 & -38.618560 & 2.4 & 0.1 & 11.0 & 0.8 & 13.46 & 0.03 & 10.34 & 0.02 & 8.76 & 0.02 \\
2 & 258.185656 & -38.617305 & 2.2 & 0.1 & 41.6 & 1.3 & 14.76 & 0.05 & 11.70 & 0.03 & 10.18 & 0.02 \\
3 & 258.219322 & -38.616643 & 5.6 & 0.1 & 37.7 & 0.3 & 14.10 & 0.03 & 10.86 & 0.02 & 9.28 & 0.02 \\
4 & 258.131277 & -38.614487 & 4.7 & 0.2 & 23.6 & 1.0 & 16.34 & 0.13 & 12.54 & 0.03 & 10.72 & 0.03 \\
5 & 258.070334 & -38.613783 & 3.4 & 0.2 & 32.2 & 1.7 & 15.59 & 0.09 & 12.43 & 0.04 & 10.96 & 0.03 \\
6 & 258.184991 & -38.614450 & 1.7 & 0.2 & 21.6 & 3.5 & 14.48 & 0.04 & 12.69 & 0.04 & 11.93 & 0.03 \\
7 & 258.072546 & -38.612958 & 1.4 & 0.3 & 21.8 & 6.3 & 15.18 & 0.06 & 13.01 & 0.04 & 12.02 & 0.04 \\
8 & 258.195536 & -38.613955 & 1.7 & 0.1 & 40.5 & 2.3 & 14.84 & 0.05 & 11.93 & 0.04 & 10.51 & 0.03 \\
9 & 258.221845 & -38.613703 & 6.7 & 0.1 & 30.5 & 0.3 & 14.36 & 0.03 & 11.28 & 0.03 & 9.82 & 0.02 \\
10 & 258.205490 & -38.613505 & 1.5 & 0.1 & 3.7 & 1.3 & 13.56 & 0.03 & 10.86 & 0.02 & 9.61 & 0.02 \\
 \noalign{\smallskip}\hline
  \end{tabular}
 \tablecomments{0.95\textwidth}{Only a portion of this table is shown here. The full version is available online and the $JK_s$-band starlight polarization data are available in https://doi.org/10.57760/sciencedb.01749 \citep{rcw120_JHK_pol}. }
\end{table*}

\begin{figure*} 
   \centering
   \includegraphics[width=0.9\textwidth]{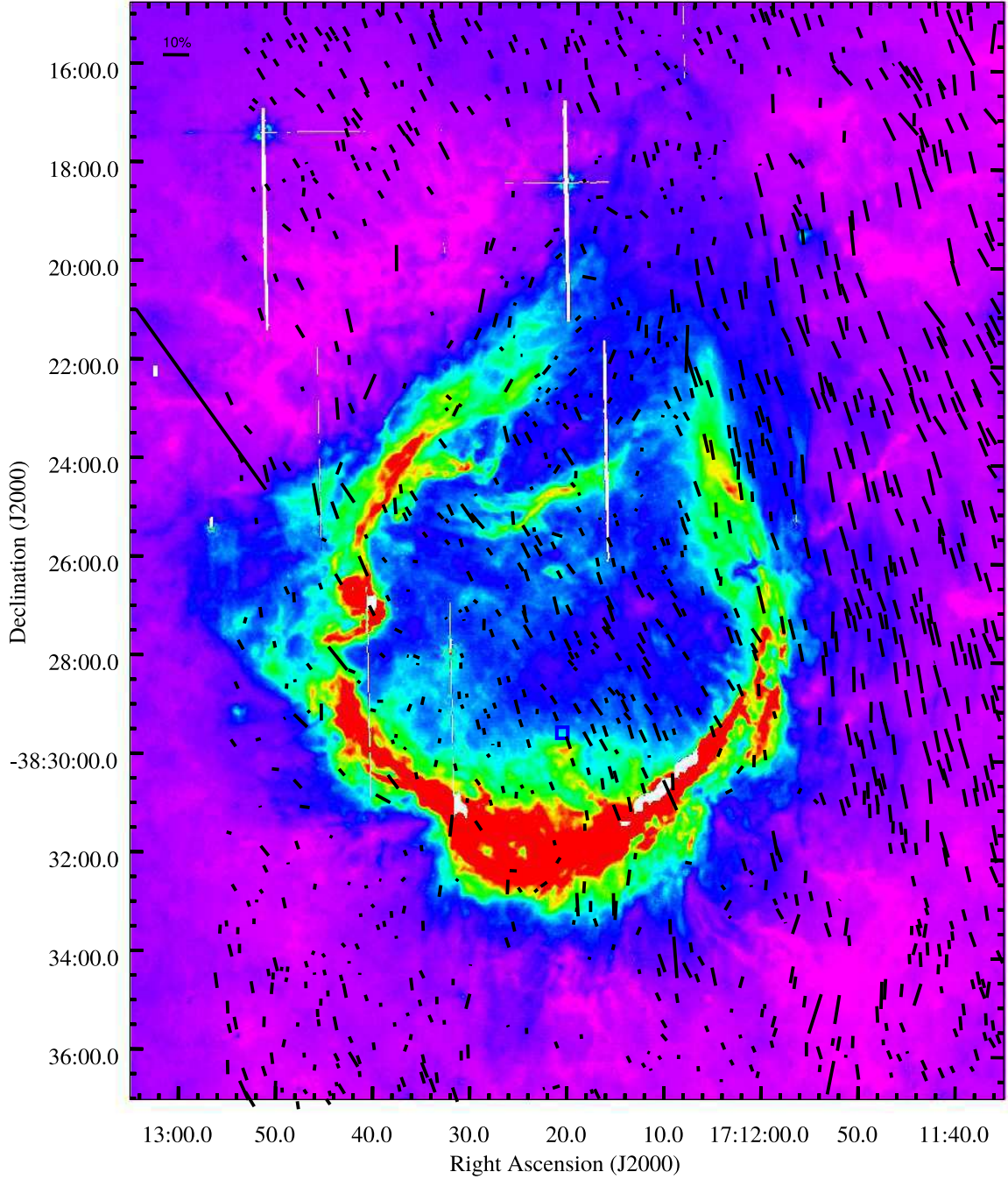}
   \caption{$H$-band starlight polarization overlaid on the Spitzer $8.0\,\mu$m image of RCW 120. The orientation of the Galactic plane is marked as the black line in the left. A 10\% segment is shown at the upper left corner.}
   \label{Fig:all_pol}
   \end{figure*}

   \begin{figure*} 
   \centering
   \includegraphics[width=0.9\textwidth]{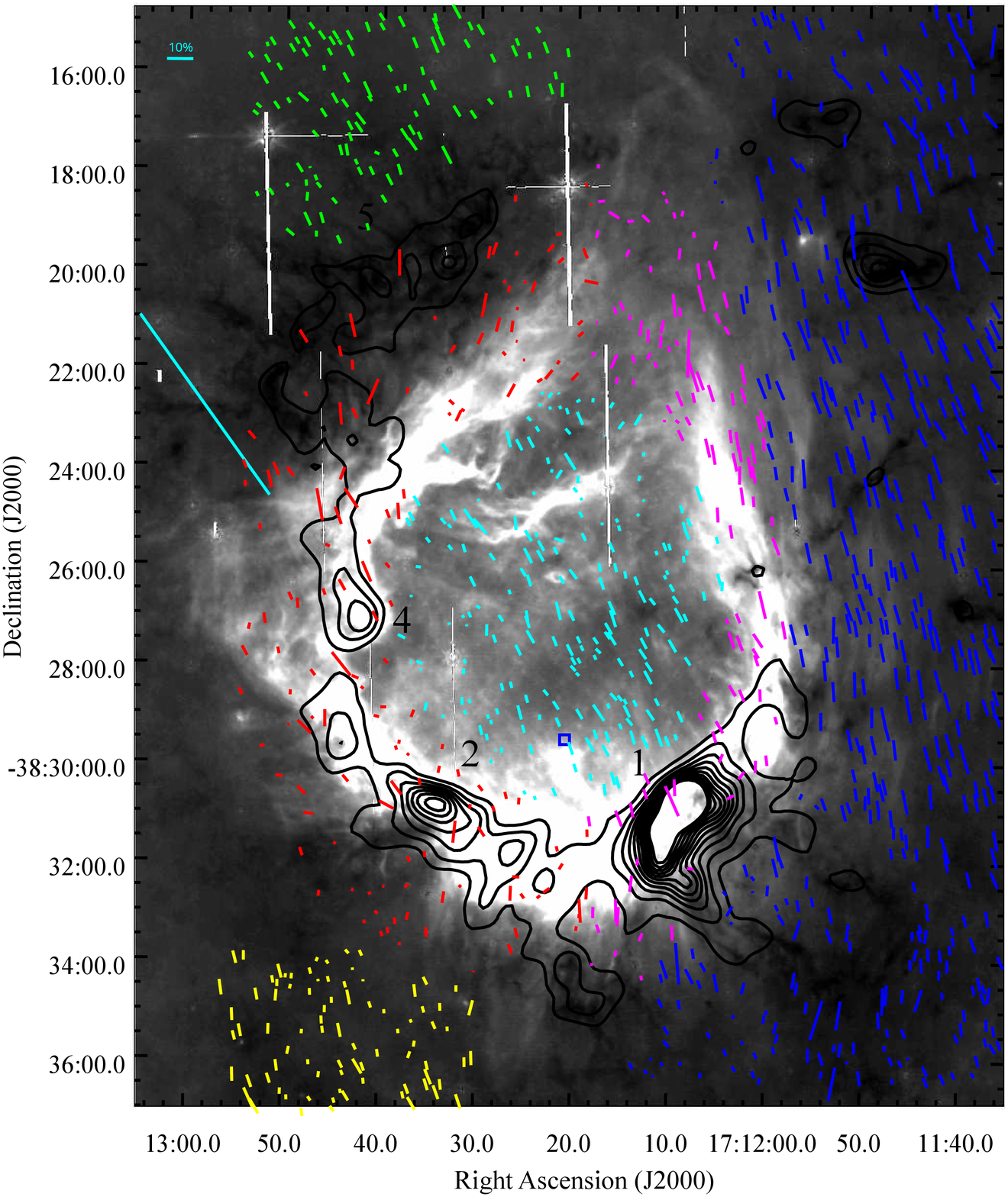}
   \caption{$H$-band starlight polarization of the subregions in six different colors overlaid on the Spitzer $8.0\,\mu$m image of RCW 120. Black contours are $870\,\mu$m dust continuum emission from the ATLASGAL survey \citep{2009A&A...504..415S}. Four dust condensations 1, 2, 4 and 5 are labeled the same as \citet{2009A&A...496..177D}. The orientation of the Galactic plane is marked as the cyan line in the left. A 10\% segment is shown at the upper left corner. } 
   \label{Fig:all_region}
   \end{figure*}
 
\subsection{The Local $B_\mathrm{sky}$ of RCW 120 }\label{Sect:onsky}

Figure~\ref{Fig:all_pol} shows the $H$-band polarization of 1143 background stars overlaid on the Spitzer $8.0\,\mu$m image. In most areas the orientations of $H$-band polarization are aligned well. The peak location between $15\degr$ and $30\degr$ of $\theta_\mathrm{PA}$, seen in Figure~\ref{Fig:all_PA}, traces the global $B_\mathrm{sky}$ of RCW 120, which is nearly, but not exactly, parallel to the Galactic plane (P.A. of $36\degr$). The difference in P.A. is about $15\degr$. This difference is also observed for the \ion{H}{ii} region N4 and its adjacent molecular cloud N4W \citep{2016ApJ...822..114C,2017ApJ...838...80C}.

The expansion of an \ion{H}{ii} region into the surrounding molecular clouds would deform the orientation of magnetic fields in areas close to the \ion{H}{ii} region \citep{2017ApJ...838...80C,2021ApJ...911...81D}. To explore in detail the influence of RCW 120 on the magnetic fields, we split RCW 120 into several subregions. The polarization segments in these subregions are signified by different colors in Figure~\ref{Fig:all_region}.\\
{\textsl Region 1.} This subregion consists of background stars in the upper left corner, shown as the green lines in Figure~\ref{Fig:all_region}. Between region 1 and the northeastern shell of RCW 120, an infrared dark cloud (IRDC) is seen in the Spitzer $8\,\mu$m image, and coincides with condensation 5 detected in the $870\,\mu$m map \citep{2009A&A...496..177D}.\\
{\textsl Region 2.} This subregion consists of the background stars lying in the west with respect to the \ion{H}{ii} region, shown as the blue lines in Figure~\ref{Fig:all_region}. \\
{\textsl Region 3.} This subregion is located in the lower left corner, far from the shell of RCW 120, shown as the yellow lines in Figure~\ref{Fig:all_region}.\\
{\textsl Eastern shell.} This subregion corresponds to the eastern shell of RCW 120, shown as the red lines. We further divide the eastern shell into the upper and lower half separated by $\mathrm{Decl.(J2000)}=-38\degr26\arcmin30\arcsec$. The orientation of the upper eastern shell is roughly along the P.A. of $135\degr$, and the orientation of the lower eastern shell is roughly along the P.A. of $40\degr$. \\
{\textsl Western shell.} This subregion represents the western shell of RCW 120, shown as the magenta lines. Similar to the division of the eastern shell, the upper western shell is along the P.A. of $15\degr$, and the lower western shell is along the P.A. of $135\degr$. \\
{\textsl Ionized zone.} This is the ionized region of RCW 120. The $H$-band polarization of background stars in this region is shown as cyan lines.

\begin{figure*} 
   \centering
   \includegraphics[width=0.99\textwidth]{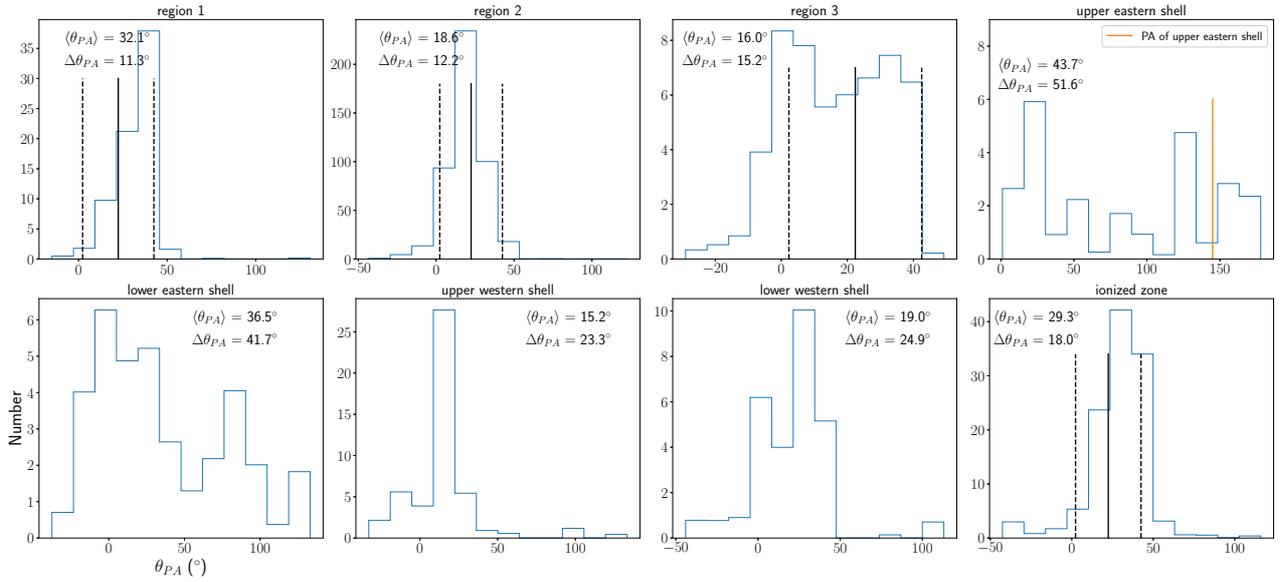}
   \caption{Histograms of $\theta_\mathrm{PA}$ in subregions. In the plots of regions 1, 2, 3 and ionized zone, the average $\langle\theta_\mathrm{PA}\rangle=22\fdg4$ and $1\sigma=20\fdg1$ deviation of the all $H$-band polarization are signified by the black solid and dashed lines, respectively. } 
   \label{Fig:PA_region}
   \end{figure*}

Figure~\ref{Fig:PA_region} features the histograms of $\theta_\mathrm{PA}$ in these subregions. The distributions of $\theta_\mathrm{PA}$ in regions 1, 2 and 3 agree well with the overall distribution of all $H$-band polarizations, and are narrow around $\langle \theta_\mathrm{PA}\rangle=22\fdg4$. Regions 1, 2 and 3 are far away from the shell of RCW 120, and are located at the different sides of RCW 120. The consistent and narrow $\theta_\mathrm{PA}$ distributions in regions 1, 2 and 3 suggest that RCW 120 has little influence on the magnetic fields in these subregions. Regions 1, 2 and 3 are representative of areas where global magnetic fields persist.

The distributions of $\theta_\mathrm{PA}$ in the eastern and western shells are different. The $\theta_\mathrm{PA}$ distributions in the eastern shell (upper and lower half) are much wider than those in regions 1, 2 and 3. In contrast, the distributions of $\theta_\mathrm{PA}$ in the western shell (upper and lower half) are consistent with those in regions 1, 2 and 3. Different from the diffuse clouds in regions 1, 2 and 3, the swept-up shell of RCW 120 is more compact and denser. The $870\,\mu$m dust continuum emission contours in Figure~\ref{Fig:all_region} outline the extinction along the shell of RCW 120. In the lower western shell where the extinction is the highest, stars with $H$-band polarization in this subregion are most likely not lying behind this part of the shell. The distributions of $\theta_\mathrm{PA}$ in the lower western shell are similar to those in regions 1, 2 and 3, indicating that the starlight polarization is tracing the global $B_\mathrm{sky}$ rather than the magnetic fields of the lower western shell, which are expected to be parallel to the shell itself. Indeed this is the limitation of the starlight polarization probe that can only work well for clouds with low to moderate extinction ($A_V\lesssim 20-30$ mag).

Because of the low extinction in the upper eastern and western shells, the detection of background stars lying behind them with starlight polarization is possible. The $\theta_\mathrm{PA}$ distributions in the upper western shell are parallel to both the orientation of the upper western shell and the global $B_\mathrm{sky}$, yielding a certain degree of ambiguity in the clarification of the magnetic fields in this subregion. A certain fraction of starlight polarization shows $\theta_\mathrm{PA}$ consistent with the orientation of the upper eastern shell, as depicted in the panel for the upper eastern shell in Figure~\ref{Fig:PA_region}. The $\theta_\mathrm{PA}$ distributions in the upper eastern shell are exclusively parallel to the orientation of the shell, indicating that the magnetic fields in this subreion are parallel to the shell orientation. The detection of magnetic fields in the denser parts of the shell is not successful. However, we do observe that the magnetic field in the low extinction parts of the swept-up shell is parallel to the shell. Magnetic fields parallel to the shell of an \ion{H}{ii} region have been observed for other \ion{H}{ii} regions similar to RCW 120, such as the N4 bubble \citep{2017ApJ...838...80C} and the S235 Main \ion{H}{ii} region \citep{2021ApJ...911...81D}. Particularly, the MHD simulations of RCW 120 exactly produced the same pattern for the $B_\mathrm{sky}$ in the shell \citep{2011MNRAS.414.1747A}. The observed pattern of $B_\mathrm{sky}$ in the upper eastern shell agrees well with both MHD simulations \citep{2007ApJ...671..518K,2011MNRAS.414.1747A} and observational results made by near-IR starlight polarization \citep{2012PASJ...64...110C,2017ApJ...838...80C,2021ApJ...911...81D} for \ion{H}{ii} regions.


The $B_\mathrm{sky}$ in the ionized zone is mostly parallel to that in regions 1, 2 and 3. The expansion of the RCW 120 \ion{H}{ii} region appears to have little effect on the magnetic fields in the ionized zone. However, this pattern of the $B_\mathrm{sky}$ in the ionized zone is largely attributed to a geometric effect. In the ionized zone, the neutral and molecular gases are dragged nearly along the LOS direction by the expansion of the \ion{H}{ii} region. The magnetic field lines are bent toward the center of the spherical \ion{H}{ii} region, and thus are largely tangential to the surface of the \ion{H}{ii} region. The on-sky deformations are very small because of this nearly LOS impact on the magnetic fields. Indeed this pattern was predicted by the MHD simulations of an \ion{H}{ii} region \citep[Figures 14, 16 and 18]{2007ApJ...671..518K}, and was described in detail for a spherical \ion{H}{ii} region \citep[Figure 7]{2001ApJ...554..916B}. The $B_\mathrm{sky}$ in the ionized zone of RCW 120 parallel to the global $B_\mathrm{sky}$ is suggested to be due to this geometric effect.

\subsection{Strength of $B_\mathrm{sky}$}
In Sect.~\ref{Sect:onsky} we describe the $B_\mathrm{sky}$ in the molecular clouds far from the \ion{H}{ii} region (regions 1, 2 and 3), in the swept-up shells which consist of neutral and molecular gas (the eastern and western shells), and in the ionized zone filled with ionized gas (the ionized zone). In this subsection we estimate the strength of $B_\mathrm{sky}$ in these subregions. For this purpose, we apply the widely used David-Chandrasekhar-Fermi (DCF) method that makes use of the starlight polarization, density and velocity dispersion of gas \citep{1951ApJ...114..206D,1953ApJ...118..113C}. The DCF method relates the strength of $B_\mathrm{sky}$ to the gas density, velocity dispersion and angular dispersion of starlight polarization by 
\begin{align}\label{equ:strength}
B_\mathrm{sky} = Q \sqrt{4\pi \rho }\frac{\delta \nu}{\delta \phi},
\end{align}
where $Q$ is the correction factor, $\rho$ is the mass density of gas, $\delta \nu$ is the one-dimensional (1D) velocity dispersion of gas and $\delta\phi$ is the angular dispersion of starlight polarization. For RCW 120, $\delta \nu$ is computed from the $^{13}$CO molecular line data described in Sect.~\ref{Sect:CO}. The gas density $\rho$ is estimated by $\rho=\mu m_\mathrm{H} N(\mathrm{H_2})/l$, where $\mu=2.83$ is the average molecular weight per hydrogen molecule \citep{2008A&A...487..993K}, $m_\mathrm{H}$ is the mass of a single hydrogen atom, $N(\mathrm{H_2})$ is molecular hydrogen column density described in Sect.~\ref{Sect:CO}, and $l$ is the thickness of RCW 120 along the LOS. The diameter of the RCW 120 bubble is about 4.5 pc \citep{2022A&A...659A..36K}. To be consistent with the flat structure of the molecular cloud associated with RCW 120 \citep{2022A&A...659A..36K}, we assume $l\sim3$ pc for the cloud associated with RCW 120 in this paper.

The $\delta \nu$ is calculated from the $^{13}$CO line emission from $-15$ to $-0.5\,\mathrm{km\,s^{-1}}$. In all the subregions, we average the $^{13}$CO line emission weighted by intensity. The averaged $^{13}$CO line emission in regions 1, 2 and 3 is fitted well by a single-peak gaussian profile. In the shell and ionized zone, the averaged $^{13}$CO line emission is blue asymmetric, indicating the influence from the expanding \ion{H}{ii} region. In regions 1, 2 and 3, the $\delta \nu$ is determined well by fitting the averaged $^{13}$CO line emission with a gaussian profile. In the shell and ionized zone, the gaussian fit to the averaged $^{13}$CO line emission fails. The $\delta \nu$ in the shell and ionized zone is the statistical mean value of the second moment map of $^{13}$CO line emission in these subregions. In the subregions where the $870\,\mu$m dust continuum emission is weak, the mean $\tau_{13}$ obtained by Equation~\ref{Equ:tao} in these subregions is $\lesssim1$, indicating optically thin $^{13}$CO line emission in these subregions. In dense subregions, such as the lower eastern and western shells, the stronger $870\,\mu$m dust continuum emission indicates that the $^{13}$CO line emission is very likely optically thick in the denser parts of the shell. The $\delta \nu$ in the lower eastern and western shells might be overestimated.

We employ two methods to estimate $\delta \phi$ from the $H$-band polarization. The statistics of the $\theta_\mathrm{PA}$ in the subregions derive a direct estimate on the angular dispersion. Figure~\ref{Fig:PA_region} displays the $\theta_\mathrm{PA}$ histograms of the subregions. We compute the average value and dispersion of $\theta_\mathrm{PA}$ weighted by $1/\delta \theta_\mathrm{PA}$. The derived angular dispersions $\Delta \theta_\mathrm{PA}$ in the subregions are tabulated in Table~\ref{tbl:pol_stats}. This method is suitable for the molecular clouds in which turbulence plays the dominant role in generating angular dispersion of starlight polarization. In fact, magnetic fields of molecular clouds may have structure due to effects such as gravitational collapse, expanding \ion{H}{ii} region or differential rotation. Consequently, this method would likely overestimate angular dispersion of starlight polarization for the clouds associated with RCW 120.

\begin{figure*} 
   \centering
   \includegraphics[width=0.99\textwidth]{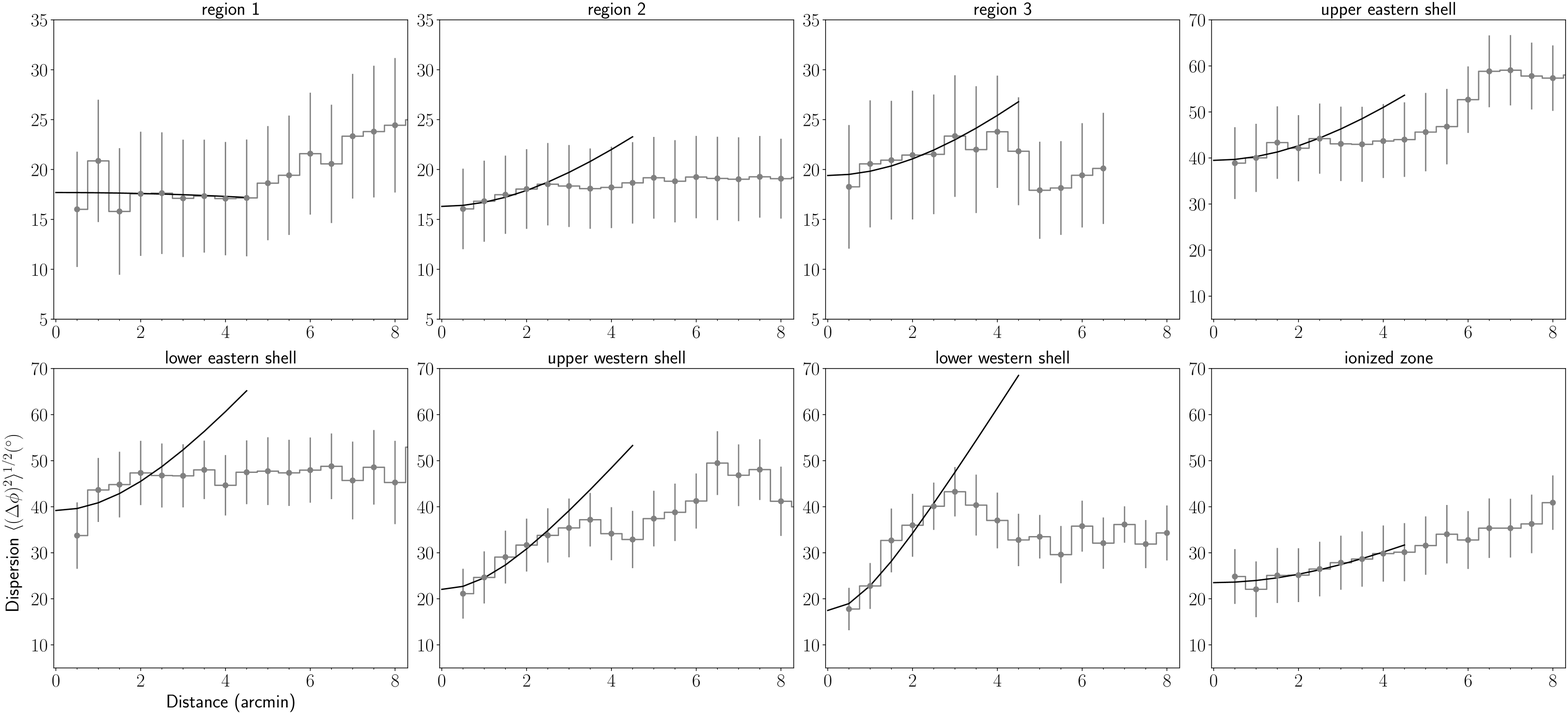}
   \caption{ADF, $\langle \Delta\phi(l)\rangle^{1/2}$, vs. displacement distance for the subregions. The turbulent contribution to the total angular dispersion is determined by the zero intercept of the fit to the data at $l=0$. The measurement uncertainties were removed prior to computing the fits to the data.} 
   \label{Fig:ADF}
   \end{figure*}

Another method, angular dispersion function (ADF), was proposed to universally estimate the angular dispersion attributed to turbulence of molecular clouds \citep{2008ApJ...679..537F,2009ApJ...696..567H,2009ApJ...706.1504H}. The ADF method obtains a measure of the difference in angle of $N(l)$ pairs of polarization segments separated by displacement $l$ through the following function

\begin{align}\label{equ:adf}
\langle \Delta\Phi^2(l)\rangle^{1/2} = \bigg \{\frac{1}{N(l)}\displaystyle\sum_{i=1}^{N(l)}[\Phi(x) - \Phi(x+l)]^2 \bigg \}^{1/2},
\end{align}
The angular dispersions attributed to the turbulent component of magnetic field $B_t(x)$ and the large scale structured field $B_0(x)$ are folded into the $\langle \Delta\Phi^2(l)\rangle^{1/2}$. Assuming that the large-scale and turbulent magnetic fields are statistically independent and the turbulent contribution to the angular dispersion is a constant $b$ for $l$ larger than the correlation length $\delta$ characterizing $B_t(x)$, the $\langle \Delta\Phi^2(l)\rangle$ approximately equals
\begin{align}\label{equ:adf1}
\langle \Delta\Phi^2(l)\rangle_\mathrm{tot} \approx b^2 + m^2l^2 + \sigma_M^2(l),
\end{align}
when $\delta<l\ll d$, where $d$ is the typical length scale of variations in $B_0(x)$, and $\sigma_M(l)$ is the measurement uncertainty. Figure~\ref{Fig:ADF} shows the ADF of the subregions along with the best fit from Equation~\ref{equ:adf1} using the first five data points ($l\leqslant2.5\arcmin$) to ensure that $l\ll d$ as much as possible. The constant $b$, characterizing the turbulent field $B_t$, is determined by the zero intercept of the fit to the ADF at $l=0$. The ratio of the turbulent to large-scale magnetic field strength is expressed as
\begin{align}
  \frac{\langle B_t^2\rangle^{1/2}}{B_0} = \frac{b}{2-b^2}.
\end{align}
We use $\delta B$ to replace ${\langle B_t^2\rangle^{1/2}}$, and define the dispersion angle $\delta \phi = \frac{\delta B}{B_0}$. Table~\ref{tbl:pol_stats} lists the $b$ and $\delta \phi$ values in degree of the subregions.

It would be interesting to compare the two dispersion angles $\Delta \theta_\mathrm{PA}$ and $\delta \phi$ for each subregions. For each subregion, $\delta \phi$ is smaller than $\Delta \theta_\mathrm{PA}$. This reduction in dispersion angle is significant for the subregions on the shell directly influenced by the expanding \ion{H}{ii}, indicating that ionization feedbacks do influence the magnetic fields on the swept-up shell. The three subregions (regions 1, 2 and 3) far from the \ion{H}{ii} region show fairly consistent dispersion angles, indicating that the turbulent magnetic field is produced by turbulence.

We use the dispersion angle $\delta \phi$ obtained from the ADF method to estimate the strength of $B_\mathrm{sky}$ in the subregions. For the subregions with small dispersion angle ($\delta \phi < 25\degr$), a correction factor $Q\approx0.5$ yields a reasonable estimate on the strength of $B_\mathrm{sky}$ \citep{2001ApJ...546..980O}. For the two subregions on the eastern shell with $\delta \phi > 25\degr$, we use the correction factor $Q\approx0.5$ as well, but the strength of $B_\mathrm{sky}$ in the two subregions is quite uncertain. From Equation~\ref{equ:strength}, we estimate the mean strength of $B_\mathrm{sky}$, and the corresponding uncertainty $\delta B$ is given by $\delta \phi = \frac{\delta B}{B_\mathrm{sky}}$. The uncertainty of field strength in the shell and ionized zone is likely the lower bound. The uncertainties of $\delta \nu$ in these subregions are comparable to $\delta \phi$. The field strength in the shell and ionized zone is very uncertain.

With the strength of $B_\mathrm{sky}$, the Alfv{\'e}nic speeds in the subregions are computed by $v_\mathrm{A} = |B_\mathrm{sky}|/\sqrt{4\pi\rho}$. The DCF method works well in the sub-Alfv{\'e}nic turbulence with Alfv{\'e}nic Mach number $M_A=\delta v/v_\mathrm{A}<1$ \citep{2008ApJ...679..537F,2021ApJ...919...79L}. Among the eight subregions, four (regions 1, 2 and 3, and ionized zone) show $M_A<1$. Moreover, the mean $\theta_\mathrm{PA}$ of the four subregions with $M_A<1$ are consistent with each other. For the global magnetic fields in the ambient cloud of RCW 120, we adopt a mean field strength in regions 1, 2 and 3, $\langle|B_\mathrm{sky}|\rangle=100\pm26\,\mu$G. Given the small dispersion angles in regions 1, 2 and 3, the field orientation is closer to being perpendicular to the LOS \citep{2008ApJ...679..537F}. In contrast, the field orientation in parts of the ionized zone is likely more parallel to the LOS, resulting from the \ion{H}{ii} region expansion along the LOS, and shows larger dispersion angle, consistent with the simulation results of field orientation more parallel to the LOS in \citet{2008ApJ...679..537F}. We suggest that the geometric projection effect due to the field orientation with respect to the LOS is small, and the total field strength $|B|\approx\langle|B_\mathrm{sky}|\rangle\approx100\pm26\,\mu$G. 



\begin{table*}

\centering
\caption{Physical Properties of the subregions}{\label{tbl:pol_stats}}
\small
 \begin{tabular}{ccccccccccc}
  \hline\hline\noalign{\smallskip}
   Region & $n_\mathrm{H_2}$ & $\delta \nu$ & $\Delta \theta_\mathrm{PA}$ & $\langle\theta_\mathrm{PA}\rangle$ & $N_\mathrm{source}$& b & $\delta \phi$  & $B_\mathrm{sky}$ & $v_\mathrm{A}$ \\
        & ($10^{3}\,\mathrm{cm^{-3}}$) & ($\mathrm{km\,s^{-1}}$) & ($\degr$) & ($\degr$) &  & ($\degr$) & ($\degr$)  & ($\mu$G) & ($\mathrm{km\,s^{-1}}$) \\
   \hline\\
   Region 1 & 1.3  & 1.6   & 11.3  & 32.1 & 128 & $17.7\pm4.0$  &$12.6\pm3.7$ &  $102\pm29$ & $3.6\pm1.1$  \\
   Region 2 & 1.3  & 1.4   & 12.2  & 18.6 & 450 & $16.3\pm1.1$  &$11.6\pm1.0$ & $98\pm22$ & $3.5\pm0.8$  \\
   Region 3 & 1.5  & 1.6   & 15.2  & 16.0 &88 & $19.4\pm1.7$ &$13.8\pm1.6$  &$98\pm27$ & $3.3\pm0.9$  \\
   Upper eastern shell & 3.3  & 1.6(0.5)  &51.6 & 43.7 & 65 &$39.5\pm3.1$  &$29.2\pm3.0$  & $72\pm44$ & $1.6\pm1.0$ \\
   Lower eastern shell  & 3.8  & 2.0(0.4)  &41.7 & 36.6 & 75  & $40.6\pm10.0$ & $30.1\pm9.3$ & $88\pm66$ & $1.9\pm1.4$ \\
   Upper western shell & 1.0 & 1.2(0.4)  & 23.3 & 15.2 &68  & $22.8\pm5.2$ & $16.3\pm4.9$& $51\pm20$ & $2.1\pm0.8 $ \\
   Lower western shell & 2.4 & 1.7(0.5)  & 24.9 & 19.0 &39 &$17.1\pm6.3$  & $12.2\pm5.9$ & $154\pm49$ & $4.1\pm1.3$\\
   Ionized zone &  1.4   & 1.8(0.4)  &18.0 & 29.3 &197 & $23.6\pm3.1$ & $16.9\pm2.9$ & $88\pm31$ &  $3.0\pm1.1$\\ 
  \noalign{\smallskip}\hline
 \end{tabular}
\tablecomments{0.86\textwidth}{In the column of $\delta \nu$, the values in parentheses are the standard deviations of $\delta \nu$ in the relevant subregions.    }
\end{table*}

\section{Discussion}
\subsection{Understanding the Morphology of RCW 120 under the View of the Magnetic Field }


The CO gas distributions are not uniform in density. The $N(\mathrm{H_2})$ of the molecular gas to the east of RCW 120 is, at least, twice that to the west of RCW120. To the north of RCW 120, where the shell breaks out, the molecular gas is little there, characterized by a mean $N(\mathrm{H_2})$ of $5.0\times10^{21}\,\mathrm{cm^{-2}}$. To the south of RCW 120, the molecular gas shows $N(\mathrm{H_2})$ comparable to the east. To first order, the expansion of the RCW 120 \ion{H}{ii} region is along the density gradient of the molecular gas, i.e., the \ion{H}{ii} region is elongated far to the north where the gas density is lower. This is consistent with the prediction of hydrodynamic simulations of RCW 120 \citep{2015MNRAS.452.2794W}. In these simulations, RCW 120 is elongated toward the northeast, where the column density is initially low. Thus the ionized gas could break through along the direction of low density \citep{2014A&A...566A..75O}. Alternative hydrodynamic simulations including stellar winds and stellar motions of RCW 120 relative to the ambient cloud, however, favor an inclination angle of $72\degr$ for RCW 120 to match the closer spherical dust shell in the south with respect to the ionizing source and the observed elongation to the northwest \citep{2019A&A...631A.170R}.

\begin{figure*} 
   \centering
   \includegraphics[width=0.98\textwidth]{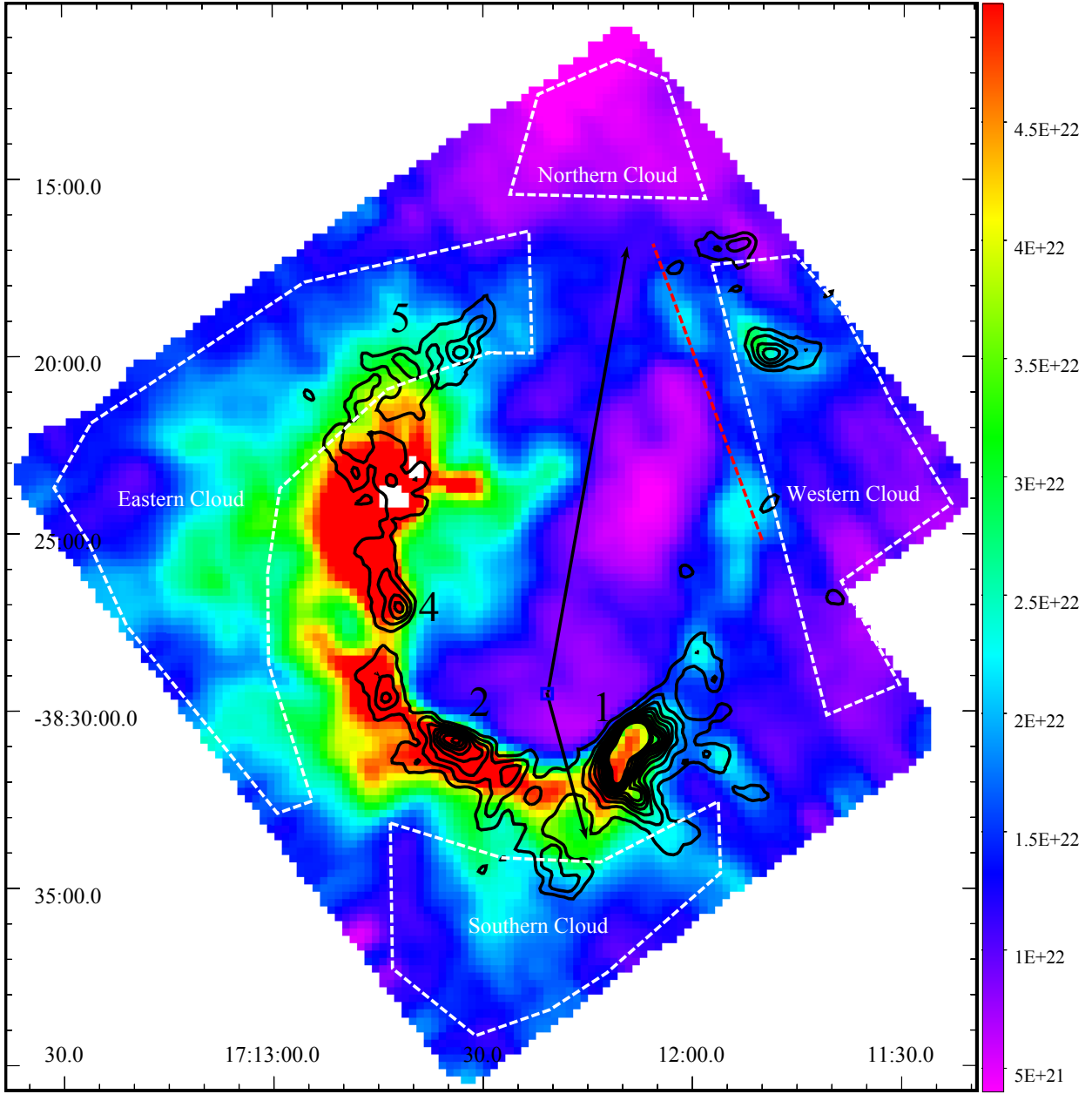}
   \caption{Hydrogen column density map derived from the $^{13}$CO line emission of RCW 120. Black contours are $870\,\mu$m dust continuum emission from the ATLASGAL survey \citep{2009A&A...504..415S}. Four dust condensations 1, 2, 4 and 5 are labeled the same as \citet{2009A&A...496..177D}. The four clouds are outlined by the white dashed lines. } 
   \label{Fig:mor}
 \end{figure*}

The above hydrodynamic simulations miss the potential impacts of the magnetic field on the asymmetric morphology of RCW 120. With the $B_\mathrm{sky}$ revealed for the first time for RCW 120, the influence of the magnetic field on the morphology of this \ion{H}{ii} region becomes invaluable. The global direction of $B_\mathrm{sky}$ in the ambient clouds is nearly parallel to the Galactic plane. The same direction is assumed when RCW 120 started to expand into the surrounding ambient clouds. Because the ionized gas can only move along the magnetic fields, the expected elongation of RCW 120 should be roughly parallel to the Galactic plane when the magnetic field is important \citep{2007ApJ...671..518K,2011MNRAS.414.1747A}. However, the observed elongation of RCW 120 is along the direction of P.A.=$-11\degr$, in contrast to the direction of the global $B_\mathrm{sky}$ which is $\sim20\degr$. Obviously, the elongation of RCW 120 is regulated not only by the magnetic field. As mentioned earlier, the density gradient from east to west and from south to north is observed in the ambient clouds of RCW 120. We divide the ambient clouds into four parts, the eastern cloud, western cloud, southern cloud and northern cloud. The four parts are illustrated in Figure~\ref{Fig:mor}. The areas adjacent to the shell are likely compressed by the expansion of the \ion{H}{ii} region, thus these areas are not included in the four parts.

\begin{figure*} 
   \centering
   \includegraphics[width=0.9\textwidth]{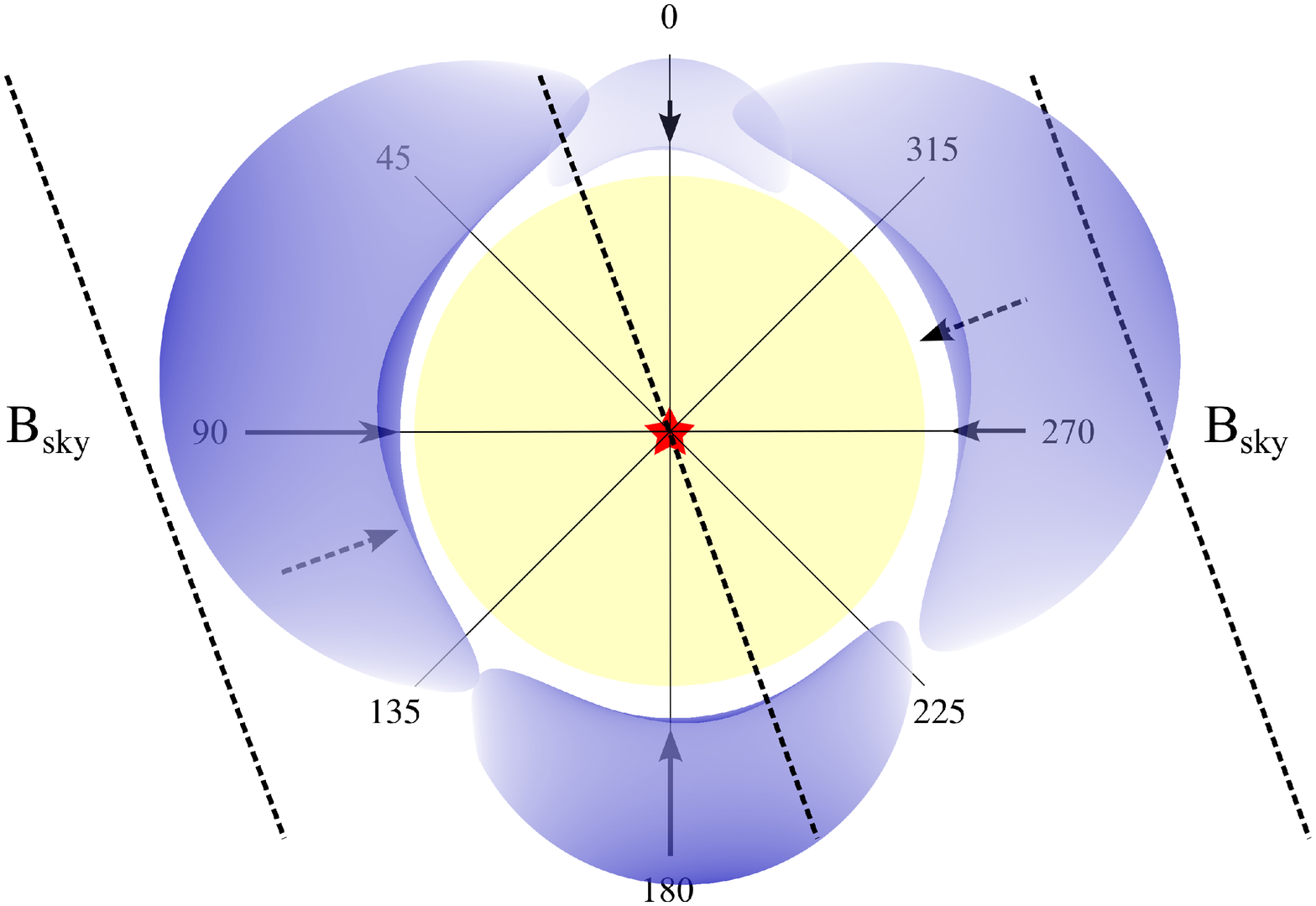}
   \caption{Diagram of the RCW 120 region when it just started to expand into the surrounding clouds.  } 
   \label{Fig:cartoon}
   \end{figure*}

For the four clouds, the average $N(\mathrm{H_2})$, $n(\mathrm{H_2})$, 1D $\delta v$, $T_\mathrm{ex}$, and the sound speed $c_s$ are computed from the CO molecular line data. For all the four clouds, the 1D $\delta v$ is much higher than $c_s$. The external dynamical pressure is computed as $P_\mathrm{dyn}/k_\mathrm{B}=\rho\,\delta v^2$ and $\rho = 2.8\,m_p\,n(\mathrm{H_2})$.  The magnetic pressure $P_\mathrm{B}/k_\mathrm{B}$ is computed by assuming the magnetic strength $|B_\mathrm{sky}|=100\,\mu$G, and it is the same for all the four clouds. The external dynamical pressure and magnetic pressure of the four clouds are tabulated in the last two columns in Table~\ref{tbl:clouds}. The four clouds have different $P_\mathrm{dyn}/k_\mathrm{B}$, depending on the $n(\mathrm{H_2})$ and 1D $\delta v$. The $P_\mathrm{dyn}/k_\mathrm{B}$ of the four clouds can be assumed to be isotropic for the expanding \ion{H}{ii} region, but the expansion of the \ion{H}{ii} region will feel strong resistance from $P_\mathrm{B}/k_\mathrm{B}$ when the expansion direction is perpendicular to the magnetic field. A sketch of the toy model for an initially spherical \ion{H}{ii} region expanding into the turbulent and magnetized ambient cloud is drawn in Figure~\ref{Fig:cartoon}.

The combined external pressure, consisting of the dynamical and magnetic pressure, is simply defined as
$$P_\mathrm{tot} = P_\mathrm{dyn} + P_B \times \cos(\theta),$$
where $\theta$ is the angle between a specific direction and the normal direction to the magnetic field. The magnetic field is along the direction P.A.$=20\degr$. When $\theta=90\degr$, the expanding direction is along the magnetic field. The \ion{H}{ii} region's expansion is largely resisted by the magnetic field when $\theta=0\degr$. The thermal pressure of the \ion{H}{ii} region is $P_i/k_\mathrm{B}=2nT$. We assume number density $n=1000\,\mathrm{cm^{-3}}$ and temperature $T=8000\,\mathrm{K}$ for RCW 120. At the earliest phase of expansion, the $P_i/k_\mathrm{B}$ of RCW 120 is much larger than the $P_\mathrm{tot}/k_\mathrm{B}$, and thus drives the \ion{H}{ii} region to expand into the ambient cloud.

\begin{table*}
  \centering
\caption{Physical Properties of the Four Clouds surrounding RCW 120}{\label{tbl:clouds}}
\small
 \begin{tabular}{ccccccccc}
  \hline\hline\noalign{\smallskip}
   Cloud & $N(\mathrm{H_2})$ & $n(\mathrm{H_2})$ & $\delta v$ & $T_\mathrm{ex}$ & $c_s$ & $v_\mathrm{A}$ & $P_\mathrm{dyn}/k_\mathrm{B}$ & $P_\mathrm{B}/k_\mathrm{B}$ \\
        &   ($10^{22}\,\mathrm{cm^{-2}}$)       & ($10^{3}\,\mathrm{cm^{-3}}$) & ($\mathrm{km\,s^{-1}}$) & (K) & ($\mathrm{km\,s^{-1}}$) & ($\mathrm{km\,s^{-1}}$) & ($10^6\,\mathrm{K\,\mathrm{cm^{-3}}}$) & $10^6\,\mathrm{K\,\mathrm{cm^{-3}}}$\\
   \hline\\
   Eastern & 2.1 & 2.2 & 1.8 & 18.1 & 0.23 & 2.7 & 2.4 & 2.9\\
   Southern  & 1.6 &1.7 & 1.9 & 16.8  & 0.22 & 3.1 & 2.1 & 2.9 \\
   Western  & 1.2 & 1.3 & 1.4 & 14.2   & 0.20 & 3.6 & 0.9 & 2.9 \\
   Northern & 0.5 & 0.6 & 1.5 & 14.1 & 0.20 & 5.3 & 0.5 & 2.9\\
  \noalign{\smallskip}\hline
\end{tabular}
\end{table*}

When the \ion{H}{ii} region just started to expand due to the thermal pressure, the \ion{H}{ii} region is spherical, as illustrated in Figure~\ref{Fig:cartoon}. We consider 13 directions tabulated in Table~\ref{tbl:pressure}. Along the magnetic field (P.A.$=20\degr$ or $200\degr$), the effective magnetic pressure is zero, resulting in the smallest $P_\mathrm{tot}/k_\mathrm{B}$ values along the two directions. However, the $P_\mathrm{dyn}$ along P.A.$=20\degr$ is very likely underestimated. Between the eastern cloud and the shell, a region of $N(H_\mathrm{2})$ about 3 times that of the eastern cloud exists. If this denser region was first exposed to the \ion{H}{ii} region, the $P_\mathrm{dyn}/k_\mathrm{B} = P_\mathrm{tot}/k_\mathrm{B}$ rises to $7.2\times10^6\,\mathrm{K\,\mathrm{cm^{-3}}}$. If the $P_\mathrm{dyn}$ along P.A.$=0\degr$ is equal to that of the eastern cloud (the value in parentheses), the $P_\mathrm{tot}/k_\mathrm{B}$ along P.A.$=0\degr$ increases to $3.4\times10^6\,\mathrm{K\,\mathrm{cm^{-3}}}$. For the upper half of the shell, the minimum $P_\mathrm{tot}/k_\mathrm{B}$ occurs along P.A.=$-10\degr$, where the $P_\mathrm{dyn}/k_\mathrm{B}$ is lowest and effective $P_\mathrm{B}$ is moderate. In the lower half of the shell, $P_\mathrm{tot}/k_\mathrm{B}$ reaches its minimum along P.A.$=200\degr$, parallel to the magnetic field.

Due to the differential $P_\mathrm{tot}$ along various directions, the expansion velocity of the \ion{H}{ii} is faster along the directions of lower $P_\mathrm{tot}/k_\mathrm{B}$, and is slower along the directions of higher $P_\mathrm{tot}/k_\mathrm{B}$. The expansion velocity also depends on the density of the ambient cloud. Let $\Delta P_\mathrm{south}/k_\mathrm{B}$ and $\Delta P_\mathrm{north}/k_\mathrm{B}$ be the net pressure of the \ion{H}{ii} region to the south and north respectively, which push the ambient molecular gas outward as the following
   $$\Delta P / k_\mathrm{B}\propto n(\mathrm{H}_2) v^2, $$
where $n(\mathrm{H_2})$ is the number density of the ambient molecular gas and $v$ is the expansion velocity. The highest $\Delta P_\mathrm{north}/k_\mathrm{B}=1.4\times10^7\,\mathrm{K\,\mathrm{cm^{-3}}}$ is along P.A.$=-10\degr$, and the highest $\Delta P_\mathrm{south}/k_\mathrm{B}=1.4\times10^7\,\mathrm{K\,\mathrm{cm^{-3}}}$ is along P.A.$=200\degr$. The expansion velocity along the two directions obeys the relation $$v_\mathrm{north}/v_\mathrm{south} \propto \left(\frac{\Delta P_\mathrm{north}}{\Delta P_\mathrm{south}}\frac{n(\mathrm{H}_2)_\mathrm{south}}{n(\mathrm{H}_2)_\mathrm{north}}\right)^{0.5}.$$
Substituting the highest $\Delta P_\mathrm{north}/k_\mathrm{B}$ and $\Delta P_\mathrm{south}/k_\mathrm{B}$, and the $n(\mathrm{H_2})$ of the northern and southern clouds respectively, we get $v_\mathrm{north}/v_\mathrm{south}=1.6$. The expansion velocity to the north is 1.6 times that to the south. This anisotropic expansion of the \ion{H}{ii} region leads to the elongated shape along P.A.$=-10\degr$, which matches the observed elongation of RCW 120. This toy model is limited to explain the observed elongation, and is too simple to match the projected morphology of RCW 120. The above toy model suggests that the morphology of an \ion{H}{ii} region depends on both the density gradient and the global magnetic field. The elongation of an \ion{H}{ii} region is toward the direction along which the total external pressure of the ambient cloud is the lowest. A similar study made from the radiation-MHD simulations predicted that the ionized regions grow anisotropically, and the ionizing stars generally appear off-center of the regions \citep{2019MNRAS.487.2200Z}.


\begin{table*}
\centering
\caption{Pressures along Various Directions}{\label{tbl:pressure}}
\small
 \begin{tabular}{ccccccc}
  \hline\hline\noalign{\smallskip}
   P.A. & $\theta$ & Cloud & $P_\mathrm{dyn}/k_\mathrm{B}$ & $\cos(\theta)\,P_B/k_\mathrm{B}$ & $P_\mathrm{tot}/k_\mathrm{B}$ & $P_i/k_\mathrm{B}$  \\
        &          &       &  ($10^6\,\mathrm{K\,\mathrm{cm^{-3}}}$) & ($10^6\,\mathrm{K\,\mathrm{cm^{-3}}}$) & ($10^6\,\mathrm{K\,\mathrm{cm^{-3}}}$) & ($10^6\,\mathrm{K\,\mathrm{cm^{-3}}}$) \\
   \hline\\
   $-10\degr$ & $60\degr$ & Northern         & 0.5 & 1.4 & 1.9 & 16 \\
   $0\degr$ & $70\degr$ & Northern (Eastern)  & 0.5(2.4) & 1.0 & 1.5(3.4)  & 16 \\
   $20\degr$ & $90\degr$ & Eastern           & 2.4(7.2) & 0   & 2.4(7.2) & 16  \\
   $30\degr$ & $80\degr$ & Eastern           & 2.4(7.2) & 0.5 & 2.9(7.7) & 16 \\
   $45\degr$ & $65\degr$ & Eastern           & 2.4(7.2) & 1.2 & 3.7(8.4) & 16   \\
   $90\degr$ & $20\degr$ & Eastern           & 2.4(7.2) & 2.7 & 5.1(9.9) & 16   \\
   $135\degr$ & $25\degr$ & Eastern          & 2.4(7.2) & 2.6 & 5.0(9.8) & 16  \\
   $180\degr$ &$70\degr$ & Southern          & 2.1 & 0.5 & 2.6 & 16 \\
   $200\degr$ & $90\degr$ & Southern         & 2.1 & 0   & 2.1 & 16 \\
   $225\degr$ &$65\degr$  & Southern         & 2.1 & 1.2 & 3.3 & 16   \\
   $270\degr$ & $20\degr$ & Western          & 0.9 & 2.7 & 3.6 & 16  \\
   $315\degr$ & $25\degr$ & Western          & 0.9 & 2.6 & 3.5 & 16  \\
   $330\degr$ & $40\degr$ & Western          & 0.9 & 2.2 & 3.1 &  16\\ 
  \noalign{\smallskip}\hline
\end{tabular}
\end{table*}

\subsection{The Dynamical Age of RCW 120 }
Both \citet{2007ApJ...671..518K} and \citet{2011MNRAS.414.1747A} proposed a critical radius or time for an \ion{H}{ii} region when the magnetic fields become significant. The magnetically critical radius is $$R_\mathrm{m}=\left( \frac{c_i}{v_\mathrm{A}}\right)^{4/3} R_0,$$ where $R_0$ is the initial Str{\"o}mgren radius in a non-magnetized medium of the same uniform density.

\citet{2011MNRAS.414.1747A} adopted an ionizing photon rate $N_\mathrm{LyC}=10^{48.5}\,\mathrm{s}^{-1}$ for the ionizing star of RCW 120, the number density $n_0=10^3\,\mathrm{cm^{-3}}$, and $R_0=0.45\,\mathrm{pc}$. These physical quantities are also adopted in this work. The sound speed, $c_i=9.8\,\mathrm{km\,s^{-1}}$, of the ionized gas is used the same way as in \citet{2011MNRAS.414.1747A}, but a higher Alfv{\'e}nic speed, $v_\mathrm{A}=3.6\,\mathrm{km\,s^{-1}}$, of the western cloud is assumed. In this work, the magnetically critical radius of RCW 120 is $R_m=1.7\,\mathrm{pc}$, which is much smaller than that of 5.7 pc in \citet{2011MNRAS.414.1747A}. This is because the observed field strength of RCW 120 ($\approx100\,\mu$G) is much stronger than that of $24.16\,\mu$G assumed in \citet{2011MNRAS.414.1747A}.  

The radius from the ionizing star to the PDR in the west (perpendicular to magnetic field) is $\sim2.0$ pc when the distance of RCW 120 is 1.68 kpc. The observed radius of RCW 120 is slightly larger than the $R_m$, indicating that magnetic effect is important for RCW 120. The dynamical age of RCW 120 is larger than the magnetically critical time $$t_\mathrm{m}=\frac{4}{7}\left( \frac{c_i}{v_\mathrm{A}}\right)^{7/3}t_0=0.53\,\mathrm{Myr},$$ where $t_0$ is the sound-crossing time of the initial Str{\"o}mgren radius of $R_0=0.45\,\mathrm{pc}$. To our knowledge, analytical solution for the MHD evolution of an \ion{H}{ii} region is currently not available. We try the approach of MHD simulations to estimate the dynamical age of RCW 120, assuming RCW 120 is expanding into a magnetized cloud of uniform density. \citet{2007ApJ...671..518K} simulated the radius of an \ion{H}{ii} region in the direction perpendicular to the magnetic field at a time $\gg t_m$. When the dynamical time of an \ion{H}{ii} region is $3t_m$, the radius in the direction perpendicular to the magnetic field is $\approx1.2R_m$ \citep[Figure 18]{2007ApJ...671..518K}. The evolution of the minimum radius perpendicular to magnetic fields of an \ion{H}{ii} region extends to evolution time $\approx10t_m$, shown as Figure 5 in the MHD simulations \citep{2011MNRAS.414.1747A}, in which $R_m=2.8$ pc and $t_m=0.61\,\mathrm{Myr}$. The ratio of observed radius of RCW 120 (2.0 pc) to the $R_m=1.7$ pc is 1.17, which yields an expected minimum radius of 3.3 pc ($1.17\times 2.8$ pc). The expected minimum radius of 3.3 pc returns an evolution time $\sim1.7\,\mathrm{Myr}$ or $\sim 3 t_m$ in Figure 5 of \citet{2011MNRAS.414.1747A}. Both MHD simulations \citep{2007ApJ...671..518K,2011MNRAS.414.1747A} point to a similar relation between radius perpendicular to magnetic field and evolution time. The observed radius of RCW 120, compared with the $R_m$ for RCW 120, suggests an MHD dynamical age $\sim3 t_m=1.6\,\mathrm{Myr}$ based on the relevant MHD simulations. 

Table~\ref{tbl:age} compiles seven dynamical ages of RCW 120 estimated by various works. Most of the studies are based on hydrodynamic simulation, whereas we and \citet{2011MNRAS.414.1747A} include the influence of the magnetic fields. The magnetic field considered in \citet{2011MNRAS.414.1747A} is much weaker than the field strength in our work. \citet{2011MNRAS.414.1747A} derived a dynamical age much younger than in our work. Because the resistance of the magnetic fields onto the expansion is not considered, the hydrodynamic age is expected to be younger than the MHD age. The most recent estimate on the hydrodynamic age of RCW 120 is in the range 0.6--1.2 Myr, considering the possible range of gas density of the ambient cloud \citep{2020A&A...639A..93F}. If we adopt the gas density ($n_0 = 2.6\times10^3\,\mathrm{cm^{-3}}$) of the western cloud in the RCW 120 \ion{H}{ii} region for Figure 11 in \citet{2020A&A...639A..93F}, the hydrodynamic age of RCW 120 is $\sim1.1\,\mathrm{Myr}$. \citet{2020A&A...639A..93F} assumed an ionizing photon rate lower than the value in our work. This hydrodynamic age of RCW 120 might be even younger if a higher ionizing photon rate is assumed. In the turbulent and magnetized molecular clouds, the apparent size of an \ion{H}{ii} region would lead to an underestimated age if the influence of the magnetic fields is not considered. When the magnetic field is weak (field strength about $10-20\,\mu$G), the size of the \ion{H}{ii} region depends more on the turbulent level or the density profile of the ambient cloud than on the magnetic field \citep{2014A&A...568A...4T,2015MNRAS.454.4484G}, therefore the influence of the weak magnetic field on the dynamical age is small. When the magnetic field of the ambient cloud is strong (field strength $\gtrsim100\,\mu$G), the dynamical age estimated by the hydrodynamic process is possibly much younger than value yielded by the MHD process. For RCW 120, the MHD dynamical age is $\sim1.6\,\mathrm{Myr}$ when field strength is $100\,\mu$G. Considering the uncertainty of measurement, the lower bound of $74\,\mu$G yields an MHD dynamical age $\sim1\,\mathrm{Myr}$ for RCW 120. The stronger the magnetic field is, the older the dynamical age of an \ion{H}{ii} region is needed to reach the observed radius. The dynamical age of RCW 120 estimated in this work is most likely older than the hydrodynamic age in \citet{2020A&A...639A..93F}. Given the uncertainty of field strength and gas properties, both ages are still comparable to each other. This is also consistent with the stage of the ionizing star of RCW 120, which is still during its main sequence stage younger than several Myr. The real age of an \ion{H}{ii} region does matter in understanding the star formation triggered by ionization feedback. If the dynamical age of an \ion{H}{ii} region is largely underestimated, then the ages of the observed YSOs lying on the swept-up shell of the \ion{H}{ii} region would likely exceed the underestimated dynamical age. This age problem conflicts with the triggered star formation mechanism. An \ion{H}{ii} region should be old enough to enable the next generation star formation triggered by the same \ion{H}{ii} region \citep{2020ApJ...897...74Z}.

\begin{table*}
\centering
    \caption{Dynamical Ages of RCW 120}{\label{tbl:age}}
  \small
  \begin{tabular}{cccccccc}
    \hline\hline\noalign{\smallskip}
$B$ 	&	 $N_\mathrm{LyC}$ 	&	  $n_0$ 	&	 $t_m$  	&	 $R_m$ 	&	 Radius 	&	   Age 	&	 Reference \\
    ($\,\mu$G) 	&	 (s$^{-1}$) 	&	 ($\times10^3\,\mathrm{cm^{-3}}$) 	&	 (Myr) 	& (pc)	   &	 (pc) 	&	 (Myr) 	&  \\	
    \hline\\
100	&	  $10^{48.5}$ 	&	 1	&	0.53	&	1.7	&	2.0 &	   1.6 	&	 This work \\
  ... 	&	  $10^{48.0}$   	&	 1.9--7.1 	&	... 	&	 ... 	&	1.8	&	0.6--1.2 	&	1 \\
 ... 	&	  $10^{48.0}$ 	&	 1--3 	&	...  	&	 ... 	&	1.7	&	 0.23-0.42 	&	 2 \\
 ... 	&	  $10^{48.1}$   	&	1	&	 ...  	&	 ... 	&	1.4	&	 0.26-0.63 	&	 3 \\
 ... 	&	  $10^{48.5}$   	&	 3-10  	&	 ... 	&	 ... 	&	1.5	&	 0.17-0.32 	&	4  \\
24.16	&	 $10^{48.5}$ 	&	1	&	2.2	&	5.7	&	1.75	&	0.2	&	  5 \\   
    ... 	&	  $10^{48.0}$ 	&	3	&	 ... 	&	 ...  	&	1.67	&	0.4	&	 6 \\
    \noalign{\smallskip}\hline
    
  \end{tabular}
  \tablecomments{0.86\textwidth}{1, \citet{2020A&A...639A..93F}; 2, \citet{2019MNRAS.483..352M}; 3,  \citet{2017MNRAS.469..630A}; 4, \citet{2013ARep...57..573P}, 5,\citet{2011MNRAS.414.1747A};  6, \citet{2007A&A...472..835Z} }
\end{table*}

\subsection{Triggered Star Formation on the Boarders of RCW 120}
The star formation triggered by an expanding \ion{H}{ii} region has been long interpreted as a promising way. The swept-up dense gas shell due to the compression of an \ion{H}{ii} region might be dense enough to start gravitational collapse. In the ISM with uniform density, the compression of an \ion{H}{ii} region is isotropic, thus a spherical or ring-like shell of enhanced density is formed on the boundary between the \ion{H}{ii} region and the molecular gas. When ordered magnetic fields are considered, the \ion{H}{ii} region becomes elongated along the magnetic field after a magnetically critical time. The swept-up shell in the direction along the magnetic field is denser and thicker than that in the direction perpendicular to the magnetic field \citep{2007ApJ...671..518K,2011MNRAS.414.1747A}. When the density gradient and magnetic field are both considered, the elongation of an \ion{H}{ii} region is determined by the summed up effect of both. For instance, the RCW 120 \ion{H}{ii} region is elongated toward the direction of P.A.$=-10\degr$, where the combined external pressure is the smallest. RCW 120 is the best representation of a realistic \ion{H}{ii} region expanding into a non-uniform and magnetized ISM. The role of the magnetic fields is discussed in the three situations.

In the direction perpendicular to the magnetic field: The eastern shell is thicker and denser than the western shell. The western cloud has a density lower than the eastern cloud. However the ordered strong magnetic field plays the dominant role in regulating the expansion of the \ion{H}{ii} region into the west. The density contrast of the western shell is about 2 in comparison with the western cloud. The strong magnetic field of the western cloud ($|B_\mathrm{sky}|\sim100\,\mu$G) greatly reduces the density contrast of the swept-up shell of the west side, which is in the direction perpendicular to the global $B_\mathrm{sky}$. The star formation triggered by the C\&C mechanism is unlikely to occur in the western shell. The strong magnetic field of the ambient cloud reduces the efficiency of triggered star formation for the swept-up shell in the direction perpendicular to the magnetic field \citep{2007ApJ...671..518K}. 

Possible RDI for the star formation in the eastern shell: The eastern shell has a high density of $5.2\times10^3\,\mathrm{cm^{-3}}$. Two condensations 3 and 4 in \citet{2007A&A...472..835Z} seen in the millimeter continuum emission are located at the eastern shell. Condensation 4 has two relatively evolved Herbig Ae/Be objects, thus star formation inside condensation 4 is possibly linked with the RDI mechanism \citep{2007A&A...472..835Z, 2017A&A...600A..93F,2020A&A...639A..93F}. The $\theta_\mathrm{PA}$ of the $H$-band starlight polarization has a large scatter in the lower half of the eastern shell ($\delta \phi=27\degr$). Together with a large $\delta v = 2.1\,\mathrm{km\,s^{-1}}$ of the molecular gas, the lower half of the eastern shell is pretty much turbulent. The dense core in the eastern shell is magnetically supercritical, given the core mass derived in \citet{2017A&A...600A..93F} and a strength of $100\,\mu$G for the magnetic field and core size about 0.2 pc \citep{2017ApJ...838...80C}. 


In the direction along the magnetic field: Two condensations 1 and 2 are located in the southern shell of RCW 120 \citep{2007A&A...472..835Z,2009A&A...496..177D}. \citet{2020A&A...638A...7Z} proposed RCW 120's compression as the mechanism for the formation of the dense shell in the south. Aided by the channel of the magnetic field, the $P_\mathrm{tot}/k_\mathrm{B}$ is small along P.A.$=200-215\degr$, thus the expansion velocity to the south is largest along this direction. The densest condensation 1 is lying in the directions of P.A.$=210-240\degr$, roughly along the global $B_\mathrm{sky}$. The dense southern shell of RCW 120 is consistent with the predictions of the MHD results \citep{2007ApJ...671..518K,2011MNRAS.414.1747A}. Moreover, even the dense southern shell is asymmetric, as revealed by the submillimeter $350\mu$m and $450\mu$m maps of RCW 120 \citep{2020A&A...638A...7Z}. The most massive condensation 1 is located at the western side of the southern shell, and the less massive condensation 2 is at the eastern half of the southern shell. The $P_\mathrm{tot}/k_\mathrm{B}$ along the direction to condensation 2 ($135\degr<$P.A.$<180\degr$ in Table~\ref{tbl:pressure}) is higher than that to condensation 1 (P.A.$=210-240\degr$ in Table~\ref{tbl:pressure}), and the expansion velocity to condensation 1 is higher than that to condensation 2. This difference leads to a higher density contrast in condensation 1 than in condensation 2, consistent with the observation that condensation 1 is more massive and denser than condensation 2. The star formation in condensation 1 has attracted intensive interests, because condensation 1 is the unique laboratory to verify the C\&C mechanism. Recent studies have demonstrated that massive star formation is on-going in the densest region of condensation 1 \citep{2017A&A...600A..93F,2018A&A...616L..10F,2020A&A...639A..93F, 2021MNRAS.503..633K}. \citet{2020A&A...639A..93F} found the dynamical age of RCW 120 to be longer than the fragmentation age of condensation 1, and they suggested the C\&C mechanism as the most likely mechanism for star-formation in condensation 1. Indeed the dynamical age of RCW 120 in this work is much larger than expected from the hydrodynamic case, implying that the triggered star formation has sufficient time to happen. By revealing for the first time the global $B_\mathrm{sky}$ of RCW 120, condensation 1 is lying in the direction along the magnetic field, which indeed channels the supersonic gas flow to collect sufficient gas to form a dense shell along the direction of the magnetic field. The on-going massive star formation in condensation 1 suggests that the magnetic field plays a role in the C\&C mechanism only for the direction along the magnetic field.

The observational evidence of RCW 120 between the triggered star formation and the magnetic field suggests that star formation triggered by an expanding \ion{H}{ii} via the C\&C mechanism in the magnetized ISM depends on whether the expanding direction is parallel to the magnetic field. If the expanding direction is along the magnetic field, like condensation 1 of RCW 120, the C\&C mechanism works for the swept-up dense shell that is compressed by the supersonic gas flow channeled by the magnetic field. If the expanding direction is perpendicular to the strong magnetic field, like the western shell of RCW 120, the density contrast of the swept-up shell is greatly reduced by the magnetic field, therefore the high density required by gravitational collapse is not created. Compared to the hydrodynamic simulations of an \ion{H}{ii} region expanding into the ISM with uniform density, the strong and ordered magnetic field in general reduces the efficiency of triggered star formation, which is consistent with the results of the radiation MHD simulations for the giant molecular clouds \citep{2021ApJ...911..128K}. However, triggered star formation via the C\&C mechanism could occur in the direction along the magnetic field.


\subsection{Core Formation Efficiency}
The star formation efficiency is difficult to estimate since the stellar masses of forming stars are unknown. \citet{2020A&A...639A..93F} estimated the core formation efficiency (CFE) of condensation 4 based on the mass of the cores and dust. Following \citet{2017A&A...600A..93F}, we estimate the CFE of condensations 1, 2 and 4. The masses of the cores in these condensations are according to Table 5 in \citet{2017A&A...600A..93F}, and the dust masses of these condensations are according to the second column of Table 7 in \citet{2017A&A...600A..93F}.

Table~\ref{tbl:cfe} shows the CFE of condensations 1, 2, 4 and 5 based on the results in \citet{2017A&A...600A..93F}. In condensation 1, six cores have estimated masses, while another six cores were detected but not discussed due to a lack of sufficient Herschel measurements \citep{2017A&A...600A..93F}. The six cores without estimated mass have size similar to the cores with moderate masses \citep{2017A&A...600A..93F}. We assume that the six cores without estimated mass contribute another $300\,M_\odot$ to condensation 1. Thus the CFE of condensation 1 can reach up to 32\% if the 12 detected cores are all included. The CFE of condensation 4 is very likely underestimated, because the core masses of the two Herbig Ae/Be objects A and B are only $1\,M_\odot$ \citep{2017A&A...600A..93F}. Assuming stellar mass of $5\,M_\odot$ for objects A and B, and the efficiency of 0.3 from core to star, the core masses of objects A and B are $34\,M_\odot$ in total. The CFE of condensation 4 then increases to 16\%. This value is consistent with the estimate (12 to 26\%) in \citet{2020A&A...639A..93F} for condensation 4. As a comparison, condensation 5 , which is little affected by RCW 120, has a much lower CFE of 1.7\%. 
The mean CFE (20\%) of condensations 1, 2 and 4 is much higher than condensation 5, indicating increase in the CFE related to triggering from ionization feedback. Previous studies for molecular clouds associated with \ion{H}{ii} regions have found that the filaments/clouds compressed by the adjacent \ion{H}{ii} regions have higher CFEs ($15-37\%$) than the regions receiving little feedback from the \ion{H}{ii} regions \citep{2007MNRAS.379..663M,2013MNRAS.431.1587E,2018A&A...609A..43X,2019A&A...627A..27X}. These results agree well with the mean CFE (20\%) of the two triggering mechanisms C\&C and RDI for RCW 120. 

Condensations 2 and 4 have comparable CFE, while the most massive condensation 1 has a CFE two to three times that of the former. The comparison between condensations 1 and 2 implies that the magnetic field has strong impact on the CFE in the C\&C mechanism, which is higher in direction parallel to the magnetic field than perpendicular to the field. In general the CFE in the RDI mechanism is comparable to the mean CFE in the C\&C mechanism, and is lower than the CFE in the C\&C mechanism along the magnetic field. We note a dependency of the CFE in the C\&C mechanism on the magnetic field, and a difference in the CFE between the C\&C and RDI mechanisms. However, the uncertainties of the clump and core masses are not included in the calculations. We wish to address this interesting topic more clearly based on more detailed studies for a number of bubble-like \ion{H}{ii} regions in a future work.   
 
\begin{table*}
  
\centering
    \caption{Core Formation Efficiency of Condensations 1, 2, 4 and 5}{\label{tbl:cfe}}
  \small
  \begin{tabular}{cccccc}
    \hline\hline\noalign{\smallskip}
Condensation	&	$M_\mathrm{clump}$ 	&	 $M_\mathrm{core}$	&	 $N_\mathrm{core}$  & CFE & Comment \\
     	&	 ($M_\odot$) 	&	 ($M_\odot$) 	&	  &  & \\	
    \hline\\
1	&	  2530 	& 502(300)	& 6(6)	& 20\%(32\%)	& C\&C \\
2 	&	  540   & 61 	&	5 & 11\%    & C\&C \\
4 	&	  350 	& 23(34) 	&	5 & 7\%(16\%)     & RDI \\
5   &     1580  & 27        & 9   & 1.7\%         & IRDC\\
    \noalign{\smallskip}\hline
    
  \end{tabular}

\end{table*}




\section{Conclusions}
We conducted the near-IR $JHK_s$ polarimetric imaging observations with the 1.4 m IRSF telescope for the well-studied bubble-like \ion{H}{ii} region RCW 120, and obtained for the first time the magnetic field of this region. The main goal of this paper is to investigate the influence of the magnetic field on the evolution of the RCW 120 \ion{H}{ii} region. The global $B_\mathrm{sky}$ in this region has a field strength of $100\pm26\,\mu$G, much stronger than the values used in the relevant MHD simulations \citep{2011MNRAS.414.1747A,2015MNRAS.454.4484G}. The strong global magnetic field shows significant impacts on the morphology and triggered star formation of RCW 120.

1. The observed morphology of RCW 120 is determined both by the density gradient and the magnetic field of the ambient clouds into which the \ion{H}{ii} region was initially expanding. The elongation is toward the direction where the turbulent pressure + magnetic pressure and the gas density of the ambient cloud are minimum.

2. The strong magnetic field has large impact on estimating the dynamical age of RCW 120. The hydrodynamic estimate and the MHD estimates with weaker field strengths underestimate the dynamical age of RCW 120 to a level younger than 1 Myr. If the strong magnetic field is included, the dynamic age of RCW 120 is $\sim\,1.6\,\mathrm{Myr}$ (for field strength of $100\,\mu$G), depending on the field strength. This age is old enough to enable the triggered star formation to occur on the boarders of RCW 120.

3. In the direction perpendicular to the magnetic field, the density contrast of the western shell of RCW 120 is greatly reduced by the strong magnetic field. Along the magnetic field, the massive and densest shell in the south shows clear evidence of triggered star formation via the C\&C mechanism. The RDI mechanism might be responsible for the star formation in the eastern shell. The strong and ordered magnetic field, in general, reduces the efficiency of triggered star formation, in comparison with the purely hydrodynamic estimates. Triggered star formation via the C\&C mechanism could occur in the direction along the magnetic field.

4. The CFEs are higher in the southern and eastern shells of RCW 120 than in the IRDC far away from the \ion{H}{ii} region, suggesting increase in the CFE related to triggering from ionization feedback. The different CFEs of the three condensations in the southern and eastern shells imply higher CFE for the C\&C mechanism than for the RDI mechanism. 

\begin{acknowledgements}
The authors thank the anonymous referee for a detailed and thoughtful review. This work is supported by the National Key Research \& Development Program of China (2017YFA0402702). We acknowledge the general Grant Nos. 11903083, 12173090, U2031202, 11873093, 11873094 from the National Natural Science Foundation of China, and the science research grants from the China Manned Space Project with No. CMS-CSST-2021-B06. This paper uses observations made at the South African Astronomical Observatory (SAAO). We thank the SAAO support astronomers and previous observer of the IRSF telescope for their technical support during the observation nights at Sutherland, South Africa, and the SAAO staff members for taking care of our lives at the SAAO. This work has made use of data from the European Space Agency (ESA) mission
Gaia (\url{https://www.cosmos.esa.int/gaia}), processed by the Gaia Data Processing and Analysis Consortium (DPAC, \url{https://www.cosmos.esa.int/web/gaia/dpac/consortium}). Funding for the DPAC has been provided by national institutions, in particular the institutions participating in the Gaia Multilateral Agreement. This research has made use of NASA’s Astrophysics Data System.

\end{acknowledgements}

\bibliographystyle{raa}
\bibliography{myrefs}

\end{CJK*}
\end{document}